\documentclass[referee]{aa}  

\usepackage{graphicx}
\usepackage{txfonts}
\graphicspath{{./figures/}}
\usepackage{natbib,twoopt}
\usepackage[breaklinks=true]{hyperref} 
\hypersetup{
  colorlinks=true,   
  urlcolor=blue,     
  linkcolor=blue     
}
\bibpunct{(}{)}{;}{a}{}{,} 

\makeatletter
\newcommand{\bibnote}[2]{\global\@namedef{#1note}{#2}}
\newcommand{\biblink}[2]{\global\@namedef{#1link}{#2}}
\makeatother

\makeatletter
  \protected\def\stonyslink{%
     \def\hyper@linkstart##1##2{}\let\hyper@linkend\@empty}
  \newcommandtwoopt{\citeads}[3][][]{%
   \href{http://adsabs.harvard.edu/abs/#3}%
        {\stonyslink \citealp[#1][#2]{#3}}
   \biblink{#3}{\href{http://adsabs.harvard.edu/abs/#3}{ADS}}}
 \newcommandtwoopt{\citepads}[3][][]{%
   \href{http://adsabs.harvard.edu/abs/#3}%
        {\stonyslink \citep[#1][#2]{#3}}
   \biblink{#3}{\href{http://adsabs.harvard.edu/abs/#3}{ADS}}}
 \newcommandtwoopt{\citetads}[3][][]{%
   \href{http://adsabs.harvard.edu/abs/#3}%
        {\stonyslink \citet[#1][#2]{#3}}
  \biblink{#3}{\href{http://adsabs.harvard.edu/abs/#3}{ADS}}}
 \newcommandtwoopt{\citeyearads}[3][][]{%
   \href{http://adsabs.harvard.edu/abs/#3}%
        {\stonyslink \citeyear[#1][#2]{#3}}
   \biblink{#3}{\href{http://adsabs.harvard.edu/abs/#3}{ADS}}}
\makeatother

\usepackage{amstext}

\begin{document} 

\title{Dynamics of charged dust in the orbit  of Venus}

\titlerunning{Charged dust in the dust ring of Venus}
\authorrunning{L. Zhou, C.Lhotka, C. Gales, Y. Narita, L.-Y., Zhou}

\author{Lei Zhou\inst{1,2,3} \and Christoph Lhotka\inst{2,4,5} \and Catalin Gales\inst{6} \and Yasuhito Narita\inst{4} \and Li-Yong Zhou\inst{1,3,7}}

\institute{
School of Astronomy and Space Science, Nanjing University, 163 Xianlin Avenue, Nanjing 210046, PR China \\
\email{zhouly@nju.edu.cn} \and
Institute of Astrophysics, University of Vienna, T\"urkenschanzstrasse 17, 1180 Vienna, Austria \\\
\email{christoph.lhotka@univie.ac.at} \and
Key Laboratory of Modern Astronomy and Astrophysics in Ministry of Education, Nanjing University, Nanjing 210046, PR China \and
Space Research Institute, Austrian Academy of Sciences, Schmiedlstrasse 6, 8042, Graz, Austria \and
Department of Mathematics, University of Rome Tor Vergata, Via della Ricerca Scientifica 1, 00133 Roma, Italy \and
Faculty of Mathematics, `Al. I. Cuza' University of Iasi, Bd. Carol I., 11, 7000506 Iasi, Romania \and
Institute of Space Astronomy and Extraterrestial Exploration (NJU \& CAST), PR China
}

\date{}

 \abstract{
  We study the dynamics of co-orbital dust in the inner Solar System, that is, the role of the solar radiation pressure, the Poynting-Robertson effect, the solar wind, and the interplanetary magnetic field, on the location, width, and stability of resonant motion of charged and micron-sized dust grains situated in the 1:1 mean motion resonance with Venus. We find deviations and asymmetry between $L_4$ and $L_5$ in the locations of the libration centers and libration width caused by nongravitational effects with analytical and numerical methods. The triangular Lagrangian points become unstable when solar radiation pressure, the Poynting-Robertson effect, and solar wind drag are considered. The Lorentz force could further destabilize the orbits, especially for small dust particles. We also compare the circular and/or elliptic restricted three-body model and a more complete model that includes all planets.
 }


%

\maketitle

\section{Introduction}

Measurements of the Venera 9 and 10 spacecraft \citepads{1979P&SS...27..951K} served as first evidence for a co-orbital dust ring around Venus. These measurements were tested later by the Helios mission \citepads{2007A&A...472..335L} and were confirmed with observations from STEREO \citepads[]{2013Sci...342..960J, 2017Icar..288..172J}.  In \citetads{2019ApJ...873L..16P} the total amount of material within the dust ring was estimated to be about $1.3\times10^{13}$\,kg, which corresponds to an equivalent of an asteroid of about 2\,km in diameter (assuming a mean bulk density of 3\,g\,{cm}$^{-3}$). According to the authors, the analysis of the measurements of the STEREO spacecraft shows that the ring is spread within a region of a radial width equal 0.06\,AU and a height of 0.1\,AU. The question now is where it comes from. If the dust ring of Venus is not a temporary phenomenon (e.g., caused by comets passing the inner Solar System), it must be continuously filled up with new material either from the interplanetary medium or from the erosion of some primordial Venus Trojans \citepads{2019ApJ...873L..16P}.

The capture probability of dust in co-orbital resonance with Venus is rather low \citepads{2012LPICo1667.6201J}, but still possible \citepads{1989Natur.337..629J}, and it is also known to exist for the other planets, for example, Earth \citepads{1994Natur.369..719D} or Jupiter \citepads{2015Icar..250..249L}. It has been shown that the capture process into mean motion resonance (MMR) between a dust particle and a planet strongly depends on the order of the resonance. It is well known that capture within the orbit of the perturber is not possible but can take place for outer resonances \citepads{1993CeMDA..57..373S, 1994Icar..110..239B,1995Icar..113..403L}. Low capture probabilities are not the only problem that needs to be solved to explain the dust ring of Venus.  Capture is also known to be a phenomenon, but it is only temporal \citepads{2015Icar..250..249L, 2019CeMDA.131...49L}. The reason is found in the effect of nongravitational perturbations, that is, solar wind drag and the force of the so-called Poynting-Robertson effect \citepads{2014Icar..232..249K}: micron-sized dust particles in the Solar System experience a net drift toward the inner part of the Solar System through loss of orbital energy that leads to a shrinking of the semimajor axes. The drift is canceled in (outer or co-orbital) resonance for finite times, but because of nonlinear perturbations, the particle starts librating with increasing amplitude around the exact resonance condition, until it is removed and again starts drifting (in radial distance) toward the Sun.

We here mainly investigate the timescales of temporary capture of micron-sized dust grains that are in co-orbital resonance with Venus. We also provide a detailed analysis of the location of the exact resonance that depends, as we show, on various parameters: the size and the charge of the dust grains, the interplanetary magnetic field strength and so on. We focus in particular on some nongravitational effect that is usually neglected in this type of studies: the role of charge in the interplanetary magnetic field.  The resulting Lorenz force may strongly affect the dynamics, as has previously been shown for micron-sized particles out of resonance \citepads{2016ApJ...828...10L} and in outer MMRs with Jupiter \citepads{2019CeMDA.131...49L}. While in the former study the authors find suitable conditions under which the drift in semimajor axes of micron-sized dust grains is reduced as a result of the normal component of the interplanetary magnetic field, a reduction in capture time has been found in \citetads{2019CeMDA.131...49L} from the enhancement of chaotic regions close to resonance because of a charge. We therefore investigate the shift in equilibria and time of temporary capture in the case of co-orbital motions of charged dust grains with Venus, including the important effect of the interplanetary magnetic field.

We state the models and methods that we used in Sec.~\ref{sec:mod}. The dynamical behavior of uncharged particles in 1:1 MMR is studied both analytically and numerically in Sec.~\ref{sec:ucp}. The effect of the Lorentz force resulting from the interplanetary magnetic field is described in detail in Sec.~\ref{sec:cp}. We present a summary and the discussion of our study in Sec.~\ref{sec:con}. Further mathematical details related to the analytical study, including a more general formula for the shift in location of the co-orbital resonance caused by nongravitational effects and the second-order solution to the characteristic equation in linear stability analysis, can be found in the appendices.

\section{Model and method}
\label{sec:mod} 

In the absence of nongravitational forces, the only perturbations of the dust grain orbits are due to additional planets and the problem reduces to the standard $N$-body problem. In the case of only one perturber, that is, Venus, the problem further reduces to the restricted three-body problem, which serves as a prototype model in Solar System dynamics. Depending on the geometry of the perturber's (Venus) orbit the planar or inclined, circular or elliptic three-body problems may be defined. These models serve as the basis for our investigations \citepads{CDbook}. However, when we consider dust grain dynamics in the Solar System, nongravitational effects can usually not be neglected \citepads{1979Icar...40....1B}. The dust grain orbits are strongly affected by the conjugated effects of the solar radiation and solar wind. 
Let $\vec r$ be the position of a dust grain in the heliocentric coordinate system. Assuming the dust grain to be a spherical object of radius $R$, let $m$ be its mass, $A=\pi R^2$ the cross-sectional area exposed to sunlight, $\rho$ the mass density, and $q$ the charge. The parameters are related by $m=4\pi\rho R^3/3$ and $q=4\pi\varepsilon_0 U R$ with surface potential $U$ and electric constant $\varepsilon_0$. As we show below, the strength of the nongravitational effects is proportional to $1/R$ (obtained from the area-to-mass ratio), and $U/R^2$ (obtained from the charge-to-mass ratio), respectively. 
In this section we essentially follow the mathematical treatment in \citetads{2019CeMDA.131...49L}, where the equation of motion for a test particle is provided by Eq. 1 therein:
\begin{equation}
        m\,\ddot{\vec{r}}=\vec{F_g}+\vec{F_{s/p}}+\vec{F_L}.
\end{equation} 
Here \vec{F_g} indicates the gravity of the Sun and additional bodies; \vec{F_{s/p}} is the force induced by the interaction of the dust grain with solar radiation and solar wind, and \vec{F_L} is the Lorentz force. For a complete treatment of the model, see \citetads{2019CeMDA.131...49L}. We only report the formalism related to the nongravitational effects that we used.

\subsection{Solar radiation pressure, the Poynting-Robertson effect, and solar wind drag}
According to \citetads{2012MNRAS.421..943K,2014Icar..232..249K}, the force caused by solar radiation pressure, the Poynting-Robertson effect, and solar wind drag can be expanded up to first order in $v/c$ as
\begin{equation}\label{eqn:fsp}
\vec{F_{s/p}}=-\nabla\frac{\mu\,m\,\beta}{r}-\frac{\mu\,m\,\beta}{r^2}\left(1+s_w\right)\left[\frac{(\vec{\dot{r}} \cdot\vec{r})\,\vec{r}}{c\,r^2}+\frac{\vec{\dot{r}}}{c}\right],
\end{equation}
where $v$ is the velocity of the particle and $c$ is the speed of light, $\nabla$ is the gradient operator, $\mu$ is the mass of the Sun, $\beta$ is the ratio between the magnitudes of the solar radiation pressure and gravity, $\vec{\dot{r}}$ is the velocity vector of the particle relative to the Sun, $r$ indicates the distance to the Sun, and $s_w$ is the ratio of solar wind drag to Poynting-Robertson effect.

The first term in Eq. (\ref{eqn:fsp}), which corresponds to the solar radiation pressure, can be expressed as the gradient of a potential proportion to gravitational potential. Therefore we incorporate this term into the Kepler motion by assigning the Sun a reduced mass of $(1-\beta)\,\mu$ when we consider the solar radiation pressure for particles. Because the force induced by the solar wind is similar to that from Poynting-Robertson effect but has a different coefficient $s_w$, we denote the combined effect of them as P-R-S effect hereafter.

\subsection{Lorentz force}
The interplanetary magnetic field has an Archimedian spiral field line pattern because the solar wind plasma takes the magnetic field radially away from the Sun, and in addition, the solar rotation winds the field lines around the rotation axis. The spiral field pattern is also referred to as the Parker spiral. The magnetic field can be oriented away from the Sun (the "away" sector) or toward the Sun (the "toward" sector), depending on how the coronal field lines are configurated. In interplanetary space near the ecliptic plane, the interplanetary magnetic field typically has a four-sector structure, which comes from an ecliptic section of the curved shaped of the magnetic equator of the Sun. In response to the solar magnetic field, the interplanetary magnetic field roughly exhibits a 22-year cycle such that the magnetic polarity reverses in every solar cycle with 11 years.

The Parker spiral model \citepads{2010JGRA..11510112W,1987ApJ...315..700B} is adopted in this paper to describe the interplanetary magnetic field. We follow the simplified analytic expression in \citetads{2019CeMDA.131...49L},
\begin{equation}\label{eqn:bb}
        \vec{B}=\frac{B_0r_0^2}{r^2}\left(\frac{\vec{r}}{r}-\frac{\Omega_s}{u_{\rm sw}}\vec{g_{\tilde{z}}\times \vec{r}}\right)\tanh\left(\alpha\frac{\vec{r}\cdot\vec{g_{\tilde{z}}}}{r}\right),
\end{equation}
where $B_0$ is the background magnetic field strength given at the reference distance $r_0$, $\Omega_s$ indicates the solar rotation rate, and $u_{\rm sw}$ is the radial speed of the solar wind, $\alpha$ is a positive parameter determining the shape of the sign change function $f(\phi)=\tanh(\alpha\phi)$, and \vec{g_{\tilde{z}}} is the unit vector pointing to the rotation axis of the Sun and can be defined by
\begin{equation}
        \vec{g_{\tilde{z}}}=\sin(i_0)\left[\sin(\Omega_0)\vec{g_x}-\cos(\Omega_0)\vec{g_y}\right]+\cos(i_0)\vec{g_z},
\end{equation}
where $i_0$ is the angle between solar equator and the ecliptic plane, and $\Omega_0$ is the ecliptic angle difference between the direction of the vernal equinox and the line of nodes between the equatorial and ecliptic planes. Following \citetads{2019CeMDA.131...49L}, $\{\vec{g_x},\vec{g_y},\vec{g_z}\}$ represents the basis of heliocentric reference frame, where \vec{g_z} points to the ecliptic pole, \vec{g_x} is aligned with the direction to the vernal equinox, and \vec{g_y} is determined by the right-handed rule.

Hence in the heliocentric ecliptic reference frame, the Lorentz force $\vec{F_L}$ is \citepads{2019CeMDA.131...49L}
\begin{equation}\label{eqn:fl}
\begin{aligned}
        \vec{F_L}=&-\frac{qB_0r_0^2}{r^2}\left[\frac{1}{r}\vec{r}\times\vec{\dot{r}}+\frac{\Omega_s}{r}\vec{r}\times(\vec{r}\times\vec{g_{\tilde{z}}})+\frac{\Omega_s}{u_{\rm sw}}(\vec{r}\times\vec{g_{\tilde{z}}})\times\vec{\dot{r}}\right] \times \\
        &\tanh\left(\alpha\frac{\vec{r}\cdot\vec{g_{\tilde{z}}}}{r}\right).
\end{aligned}
\end{equation}
In addition to the position and velocity, the acceleration caused by the Lorentz force is mainly dependent on the charge-to-mass ratio $q/m$ of particles.

We note that the interplanetary magnetic field given by Eq. (\ref{eqn:bb}) does not include time-dependent effects, such as variations in the axial tilt or the solar cycle. However, it is based on the classical Parker spiral including the axial tilt of the solar dipole with respect to the ecliptic \citepads{2019AnGeo..37..299L}. The approach is therefore only valid on secular timescales, that is, where time-dependent effects due to solar activity are canceled because of averaging. For additional information, see \citetads{2019AnGeo..37..299L}.

\subsection{Parameters and numerical setup}
The values we used and the corresponding references for parameters involved with nongravitational effects are summarized in Table.~\ref{tab:param}. The orbital elements of the planets at epoch of JD 245\,1545.0 in the ecliptic system of J2000.0 are taken from the JPL HORIZONS system\footnote{\url{ssd.jpl.nasa.gov/horizons.cgi}} \citepads{1996DPS....28.2504G}. By convention, the orbital elements with respect to the ecliptic plane in the heliocentric reference frame, which are semimajor axis, eccentricity, inclination, argument of perihelion, longitude of the ascending node and mean anomaly, are denoted by $a$, $e$, $i$, $\omega$, $\Omega,$ and $M$, respectively. We use the subscript `V' to indicate Venus throughout.

\begin{table}[htbp]
                \centering
        \caption{Summary of parameters we used for nongravitational effects.}
        \begin{tabular}{lll}
        \hline\hline
        Symbol & Values & Reference \\
        \hline
        $\alpha$ & 100  & \citetads{2019CeMDA.131...49L} \\
        $B_0$    & 3 nT & \citetads{2012bsw..book.....M} \\
        $i_0$    & $7.15^\circ$ & \citetads{2005ApJ...621L.153B} \\
        $\Omega_0$ & $73.5^\circ$ & \citetads{2005ApJ...621L.153B} \\
        $\Omega_s^{-1}$ & 24.47 d & \citetads{2012bsw..book.....M} \\
        $r_0$    & 1 AU & \citetads{2012bsw..book.....M} \\
        $s_w$    & 1/3  & \citetads{2019CeMDA.131...49L} \\
        $u_{\rm sw}$  & 400 km/s  & \citetads{2012bsw..book.....M} \\
        \hline
        \end{tabular}
        \label{tab:param}
\end{table}

\begin{figure}
        \centering
    \resizebox{\hsize}{!}{\includegraphics{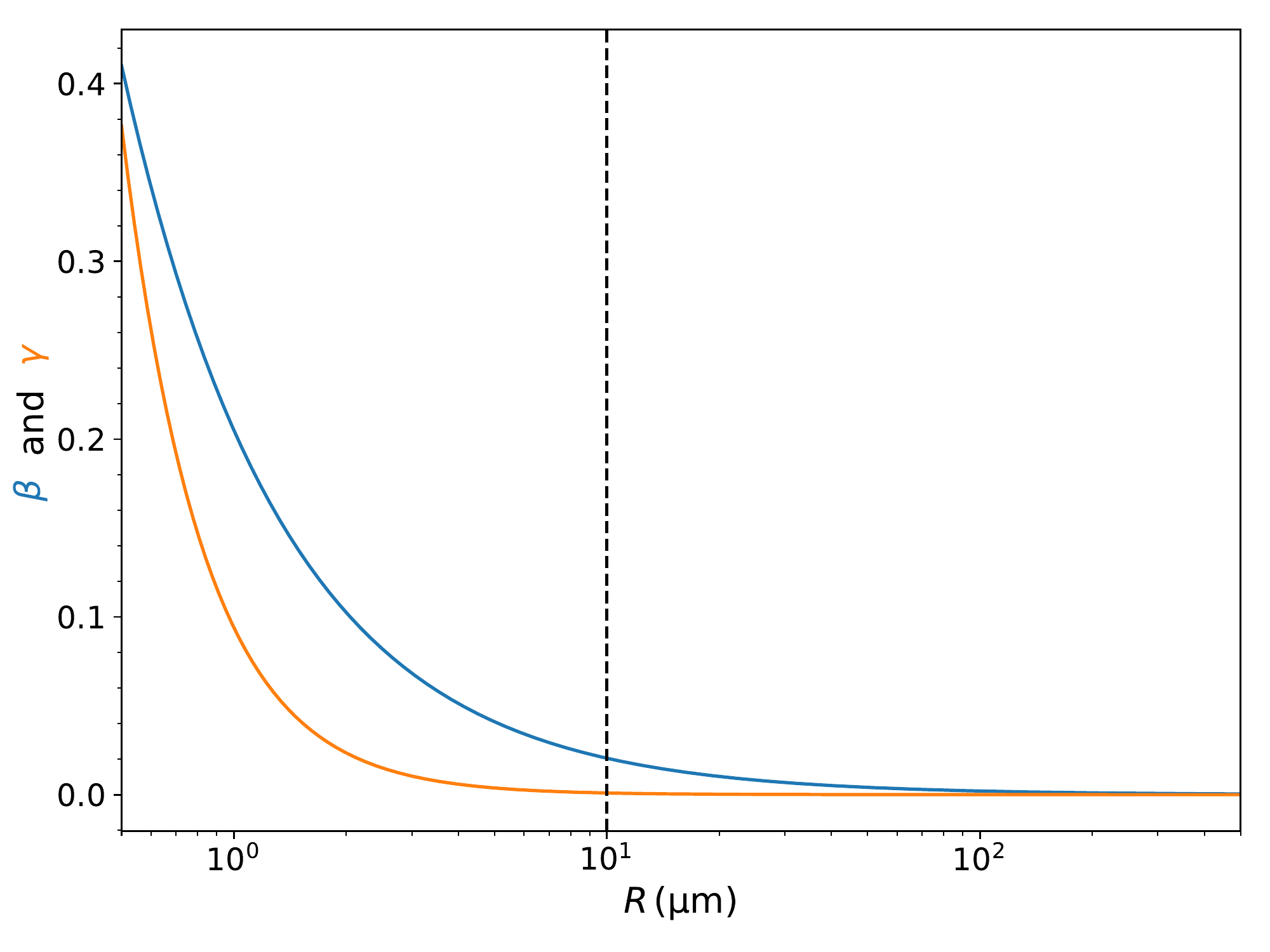}}
                \caption{Parameter $\beta$ (blue) and $\gamma$ (orange) assuming a surface potential of 10 V. The dashed line shows $R=10\,\mu{\rm m}$, for which we find $\beta=0.0205$, $\gamma=0.00094$.}
                \label{fig:rbg}
        \end{figure}

We investigated the dynamics of co-orbital particles under different magnitudes of nongravitational effects, which can be evaluated by $\beta$ and $q/m$. Setting $\rho=2.8\,{\rm g/cm^3}$, we have
\citepads{1994Icar..110..239B,2019CeMDA.131...49L}
\begin{equation}\label{eqn:bqm}
        \beta=0.205/R,\quad \gamma=0.0094U/R^2,
\end{equation}
where the radius $R$ is in microns and the surface potential $U$ is in Volt, and $\gamma$ is the numerical value of $q/m$ in C/kg. Given a surface potential of 10 V, the dependence of $\beta$ and $\gamma$ on the size of particle is depicted in Fig.~\ref{fig:rbg}. As we show in Sec.~\ref{subsec:ns}, the $\beta$ value that we consider for the Venus dust ring does not exceed 0.34, which corresponds to the minimum particle radius of 0.6 ${\rm \mu m}$. 

In numerical simulations, we modified the symplectic massive body algorithm (SyMBA) integrator \citepads{1998AJ....116.2067D,2000AJ....120.2117L} to include nongravitational effects for dust particles. SyMBA is a second-order symplectic integrator that can handle close encounters between objects and also a nonconservative effect. It is based on a variant of standard mixed variable symplectic (MVS) methods and could switch to an improved multiple time-step method when close encounters occur. Except for solar radiation pressure, which is incorporated into the gravitational potential, we included other nongravitational effects by simply adding corresponding acceleration in every time step.

In our analytical study, we considered the planar circular restricted three-body problem (CRTBP) in the barycentric rotating reference frame where the unit of mass is the total mass of the Sun and planet (Venus in this paper), and the unit of length is the semimajor axis of the planet. When $G=1$, the unit of time should be $T'/(2\pi)$, where $T'$ is the period of the mean motion of the planet. The numerical value of the speed of light $c$ in the normalized unit system is denoted by $c_v$ and can be calculated by 
\begin{equation}\label{eqn:cv}
        c_v\approx\frac{173\,T'}{2\pi\,a'}=\frac{173}{n'a'},
\end{equation}
where $a'$ and $n'$ are the semimajor axis and mean motion of the planet, and $T'$ is the corresponding period in days. For Venus, $c_v$ is 8561. 

The framework of an elliptic restricted three-body problem (ERTBP) in heliocentric reference frame is mainly adopted in our numerical simulations. Like the CRTBP, the ERTBP here also consists of the Sun, Venus, and a massless particle, but Venus is assumed to move in its real orbit with low eccentricity (0.00673) and inclination ($3.39^\circ$) in the ecliptic system of J2000.0 centered on the Sun instead of on a circular orbit. We also considered the complete model of the current planetary configuration of our Solar System based on J2000.0 in our numerical simulations to study the effect of the planetary perturbations. All planets in the Solar System are included, and the Earth and Moon are treated as a whole body in their barycenter. All numerical simulations were carried out within the framework of the ERTBP, unless otherwise specified. Although the CRTBP is based on the barycentric reference frame, the orbital elements are always defined with respect to the Sun for all models in this paper.
 
\section{Uncharged problem} \label{sec:ucp}

The triangular Lagrangian points $L_4$ and $L_5$ are dynamically stable for all planets in our Solar System in the pure gravitational CRTBP \citepads[see e.g.][]{1999ssd..book.....M}. In the vicinity of these points, an important reservoir for celestial objects such as asteroids and dust is therefore located. The investigation of these enormous populations could not only reveal the dynamical mechanisms behind the complicated orbital behavior, but also determine the possible origin of these objects and verify the scenarios proposed to describe the early stage of our Solar System.

\begin{figure}
        \centering
    \resizebox{\hsize}{!}{\includegraphics{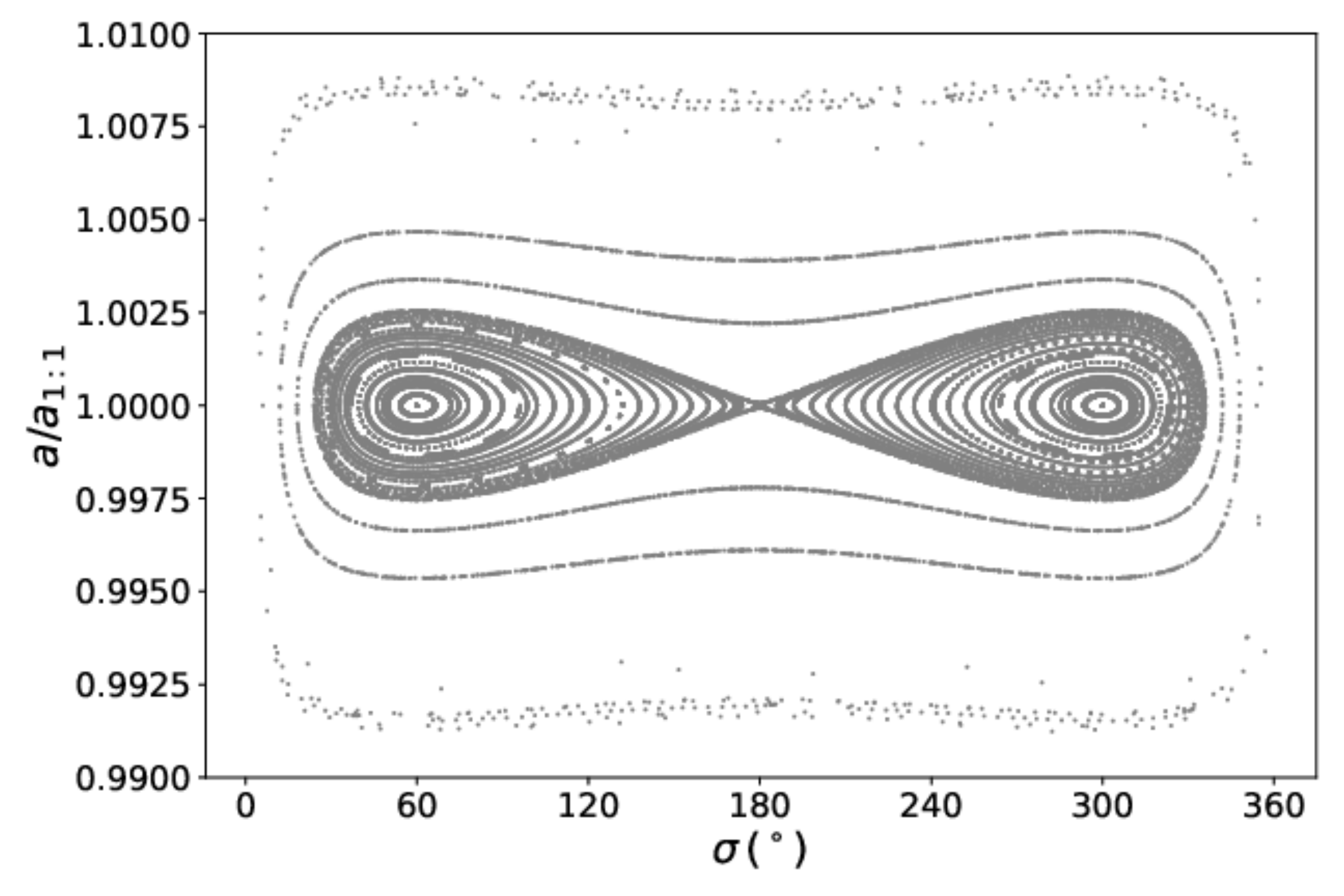}}
    \resizebox{\hsize}{!}{\includegraphics{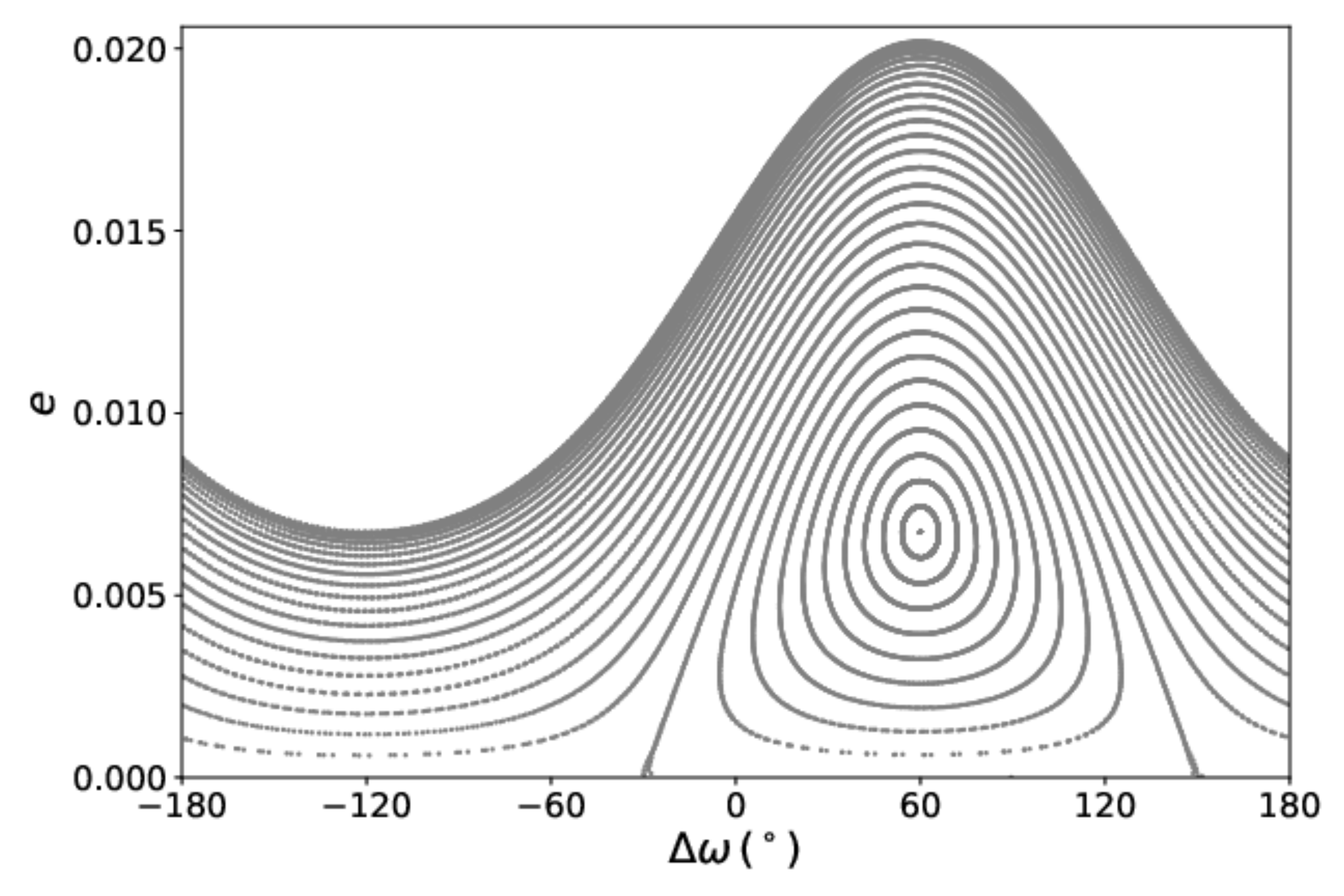}}
                \caption{Phase portrait of the 1:1 MMR with Venus in the ERTBP. The \textit{upper} panel shows the $(\sigma,a)$ plane with resonant angle $\sigma$ and (normalized) semimajor axis $a,$ while the \textit{lower} panel shows the $(\Delta\omega,e)$ plane (for $L_4$ only) with apsidal difference $\Delta\omega$ and eccentricity $e$.}
                \label{fig:phport}
        \end{figure}

The typical phase portraits of the 1:1 MMR with Venus obtained within the framework of the ERTBP on $(\sigma,a)$ and $(\Delta\omega,e)$ planes are shown in Fig.~\ref{fig:phport}. Here, $\sigma=\lambda-\lambda_{\rm V}$ is defined as the resonant angle of 1:1 MMR where $\lambda=\omega+\Omega+M$ is the mean longitude. We note for planar CRTBP that $\omega$ and $\Omega$ have no definition so that $\sigma=M-M_{\rm V}$. We use $\Delta\omega=\omega-\omega_{\rm V}$ to denote the difference in argument of perihelion between the particle and Venus, and $e$ is the eccentricity of the dust particle. We integrated 60 orbits for each plot in the framework of the ERTBP. Two stable points in the center of tadpole regimes, which are denoted by $L_4$ and $L_5$, reside symmetrically at $60^\circ$ and $300^\circ$, respectively. The saddle point $L_3$, located at $180^\circ$, separates two tadpole regimes and the horseshoe regime. We note that the semimajor axis $a$ in the plot is normalized by $a_{1:1}=a_{\rm V}\,(1-\beta)^{1/3}$, which is the location of the 1:1 MMR for dust particles (see below). Here $\beta=0$ and so $a_{1:1}=a_{\rm V}$. The apsidal difference $\Delta\omega$ could librate around the center for orbits nearby. The center $(\Delta\omega_k,e_k)$ with $k=4,5$ is $(\pm60^\circ,e_{\rm V})$ for $L_4$ and $L_5$. The phase portraits on the $(\Delta\omega,e)$ plane are also symmetric for $L_4$ and $L_5$ when $\beta=0$, so that we only show the portrait for the former in Fig.~\ref{fig:phport}.

\subsection{Location of triangular Lagrangian points}\label{subsec:loc}

The theoretical derivation in this section is based on the planar CRTBP. We aim to derive formulae to predict the shift of the Lagrangian points $L_4$ and $L_5$ due to the nongravitational effects, that is, radiation pressure and the P-R-S effect. We worked in synodic coordinates where $\vec{r_1}$ and $\vec{r_2}$ indicate the position vectors from the Sun and planet, respectively. The distance to the Sun $r$ (in Eqs. (\ref{eqn:fsp}), (\ref{eqn:bb}), and (\ref{eqn:fl})) therefore equals $r_1$. In the rotating reference frame centered on the center of mass of the Sun-planet system with the planet fixed at the $+x$ -axis, the equations of motion including $F_{s/p}$ in the planar case \citepads{1970SvA....14..176C,1980ApJ...238..337S} can be written as

\begin{equation}\label{eqn:eomx}
\begin{aligned}
\ddot{x}&-2\dot{y}+\frac{\gamma_c\,(1+s_w)}{r_1^2}\left\{\dot{x}+\left[\dot{x}\,(x+\mu')+\dot{y}\,y\right]
\frac{x+\mu'}{r_1^2}\right\}= \\
&-\frac{(1-\beta)\,\mu\,(x+\mu')}{r_1^3}-\frac{\mu'\,(x-\mu)}{r_2^3}+x+\frac{\gamma_c\,(1+s_w)\,y}{r_1^2},
\end{aligned}
\end{equation}
\begin{equation}\label{eqn:eomy}
\begin{aligned}
\ddot{y}&+2\dot{x}+\frac{\gamma_c\,(1+s_w)}{r_1^2}\left\{\dot{y}+\left[\dot{x}\,(x+\mu')+\dot{y}\,y\right]\frac{y}{r_1^2}\right\}= \\
&-\frac{(1-\beta)\,\mu\,y}{r_1^3}-\frac{\mu'\,y}{r_2^3}+y-\frac{\gamma_c\,(1+s_w)\,(x+\mu')}{r_1^2},
\end{aligned}
\end{equation}
where $\mu'=1-\mu$ is the mass of the planet, $(x, y)$ are the coordinates of the dust particle with respect to the barycenter of the Sun and Venus, $r_2$ indicates the distance to the planet, and $\gamma_c$ is a parameter given by $\gamma_c\equiv\beta\,\mu/c=\beta\,(1-\mu')/c$, which mainly determines the magnitude of the P-R-S effect for particles on specific orbits (see Eq. (\ref{eqn:fsp})).

In addition to our analytical solutions to Eqs. (\ref{eqn:eomx}) and (\ref{eqn:eomy}), some numerical simulations within the framework of the ERTBP were conducted to verify the theoretical formulae. We also included other planets in the Solar System to study the effect of the planetary perturbations on the co-orbital dynamics for Venus.

\subsubsection{Analytical study}
When $\beta\neq0$, nongravitational effects introduce additional forces that are related to the position and velocity of the particles. Therefore the equation of motion in the classic CRTBP should be modified, and the equilibrium solutions were changed accordingly. Because nongravitational effects are dependent on $\beta$, which is determined by the size of the particle, the locations of the triangular Lagrangian points rely on the size of particles.

In the rotating reference frame of the CRTBP, the locations of libration points considering solar radiation pressure and the P-R-S effect are determined by setting $\ddot{x}=\ddot{y}=\dot{x}=\dot{y}=0$ in Eqs. (\ref{eqn:eomx}) and (\ref{eqn:eomy}).
Following \citetads{1980ApJ...238..337S}, we defined the following symbols:
\begin{eqnarray}
        \label{eqn:r10}r_1^{(0)}&=&(1-\beta)^{1/3}=\delta,\\
        \label{eqn:r20}r_2^{(0)}&=&1,\\
        \label{eqn:x00}x^{(0)}&=&\frac{(1-\beta)^{2/3}}{2}-\mu'=\frac{\delta^2}{2}-\mu'=x_0,\\
        \label{eqn:y00}y^{(0)}&=&\pm(1-\beta)^{1/3}\left[1-\frac{(1-\beta)^{2/3}}{4}\right]^{1/2} \nonumber \\
        &=&\pm\delta\left(1-\frac{\delta^2}{4}\right)^{1/2}=y_0,
\end{eqnarray}
which are also the analytical solution of the locations of $L_4$ and $L_5$ when we ignore the P-R-S effect and consider only solar radiation pressure. We note that the plus$\text{}$ sign in Eq. (\ref{eqn:y00}) corresponds to $L_4$ while the $\text{minus}$ corresponds to $L_5$.

We multiplied Eq. (\ref{eqn:eomx}) by $y$ and Eq. (\ref{eqn:eomy}) by $(x+\mu')$, then subtracted them. Making use of $\ddot{x}=\ddot{y}=\dot{x}=\dot{y}=0$ and recalling that $r_1^2=(x+\mu')^2+y^2$, we can obtain
\begin{equation}\label{eqn:r2}
        r_2 = \left[1-\frac{\gamma_c\,(1+s_w)}{\mu'\,y}\right]^{-1/3}.
\end{equation}
To keep $r_2>0$ (it is always satisfied for $L_5$), we should ensure that for $L_4$
\begin{equation}
        \gamma_c<\frac{\mu'\,y}{(1+s_w)}.
\end{equation}
Or equivalently,
\begin{equation}\label{eqn:brange}
        \beta<\frac{\mu'\,y\,c}{(1+s_w)\,(1-\mu')}.
\end{equation}
Applying Eq. (\ref{eqn:brange}) to Venus and taking $y\approx{y_0}$, we can obtain $\beta<0.01356$ for $L_4$. Substituting the expression of $r_2$, given by Eq. (\ref{eqn:r2}), in the right-hand side of Eq. (\ref{eqn:eomy}) which is equal to 0, we obtain
\begin{equation}\label{eqn:r11}
        \left[\gamma_c\,(1+s_w)+(1-\mu')\,y\right]r_1^3-\gamma_c\,(1+s_w)\,(x+\mu')\,r_1=\delta^3\,(1-\mu')\,y.
\end{equation}
Since 
\begin{equation}\label{eqn:xpu}
        x+\mu'=\frac{r_1^2-r_2^2+1}{2},
\end{equation}
we can rewrite Eq. (\ref{eqn:r11}) as
\begin{equation}\label{eqn:r12}
\begin{aligned}
        &\left[\frac{\gamma_c\,(1+s_w)}{2}+(1-\mu')\,y\right]r_1^3+\frac{\gamma_c\,(1+s_w)}{2}\times \\
        &\left\{\left[1-\frac{\gamma_c\,(1+s_w)}{\mu'\,y}\right]^{-2/3}-1\right\}\,r_1=\delta^3\,(1-\mu')\,y.
\end{aligned}
\end{equation}

The $Nth^{\rm }$ order solution to Eqs. (\ref{eqn:r12}) and (\ref{eqn:r2}) can be written as a series expansion in $\gamma_c$:
\begin{equation}\label{eqn:gensol}
\begin{aligned}
        &x^{(N)}=\sum^N_{m=0}\,p^{(m)}_{4,5}(\beta,\mu')\,\gamma_c^m\,(1+s_w)^m,\\
        &y^{(N)}=\sum^N_{m=0}\,q^{(m)}_{4,5}(\beta,\mu')\,\gamma_c^m\,(1+s_w)^m,
\end{aligned}
\end{equation}
where $p^{(m)}_{4,5}(\beta,\mu')$ and $q^{(m)}_{4,5}(\beta,\mu')$ are coefficients of the $mth^{\rm }$ order term for $L_4$ or $L_5$. Obviously $p^{(0)}_4=p^{(0)}_5=x_0$ and $q^{(0)}_4=q^{(0)}_5=y_0$. Solutions up to the third order are derived in Appendix~\ref{apd:sol}. Because the solutions are complicated, we tried to simplify them by assigning $\beta=0$ in $p^{(m)}_{4,5}(\beta,\mu')$ and $q^{(m)}_{4,5}(\beta,\mu')$ for $m>0$ terms. Therefore, the coefficients $p^{(m)}_{4,5}$ and $q^{(m)}_{4,5}$ with $m>0$ are only dependent on $\mu'$. This is feasible because in the case of Venus, even the original formulae shown in Appendix~\ref{apd:sol} are well consistent with the numerical solutions to the equations of motion only when $\beta\lesssim 0.01$ for both $L_4$ and $L_5$ (see below). In this region we have small differences in the resonant angles of libration centers $\sigma_k$ ($k=4,5$) between original and simplified formulae, up to $0.12^\circ$ and $0.007^\circ$ for $L_4$ and $L_5$, respectively. Up to the third order in $\gamma_c$, we have
\begin{equation}\label{eqn:simp}
        \begin{aligned}
                p^{(1)}_4&=&-p^{(1)}_5&=&&-\frac{2-\mu'}{3 \sqrt{3} (1-\mu')\mu},& \\
                p^{(2)}_4&=&p^{(2)}_5&=&&\frac{(64-9 \mu') \mu'-44}{162 (1-\mu')^2 \mu'^2},& \\
                p^{(3)}_4&=&-p^{(3)}_5&=&&\frac{u [-3 (102-\mu') \mu'+520]-248}{729 \sqrt{3} (1-\mu')^3 \mu'^3},& \\
                q^{(1)}_4&=&q^{(1)}_5&=&&\frac{2-3 \mu'}{9 \mu'(1-\mu')},& \\
                q^{(2)}_4&=&-q^{(2)}_5&=&&\frac{-\mu'  (16 - 13 \mu')+4}{54 \sqrt{3} (1 - \mu')^2 {\mu'} ^2},& \\
                q^{(3)}_4&=&q^{(3)}_5&=&&-\frac{3 \mu'  [\mu'  (27 \mu' -58)+16]+40}{2187 (1 - \mu')^3 {\mu'} ^3}&.
        \end{aligned}
\end{equation}

\begin{figure}
\centering
\resizebox{\hsize}{!}{\includegraphics{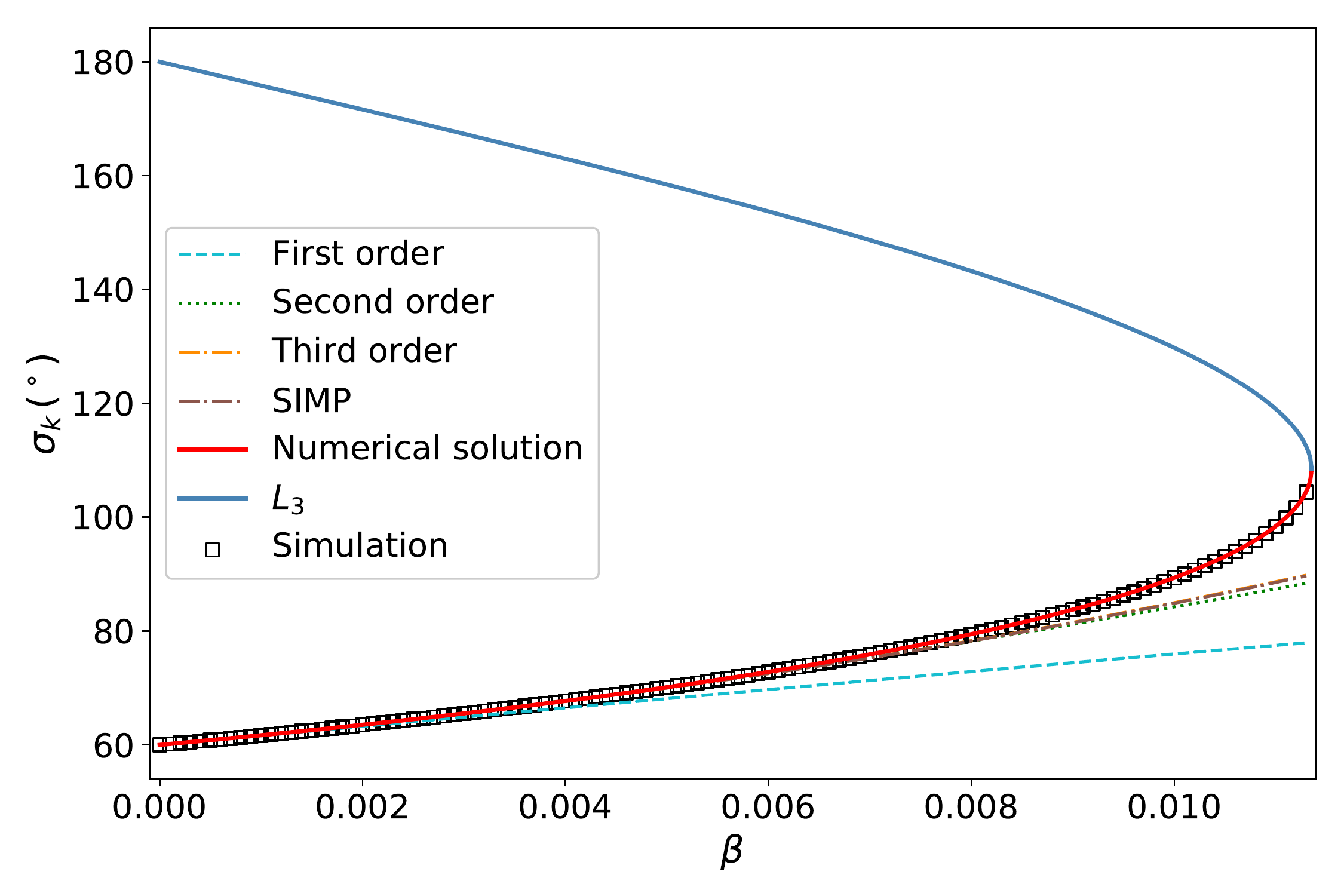}}
\resizebox{\hsize}{!}
{\includegraphics{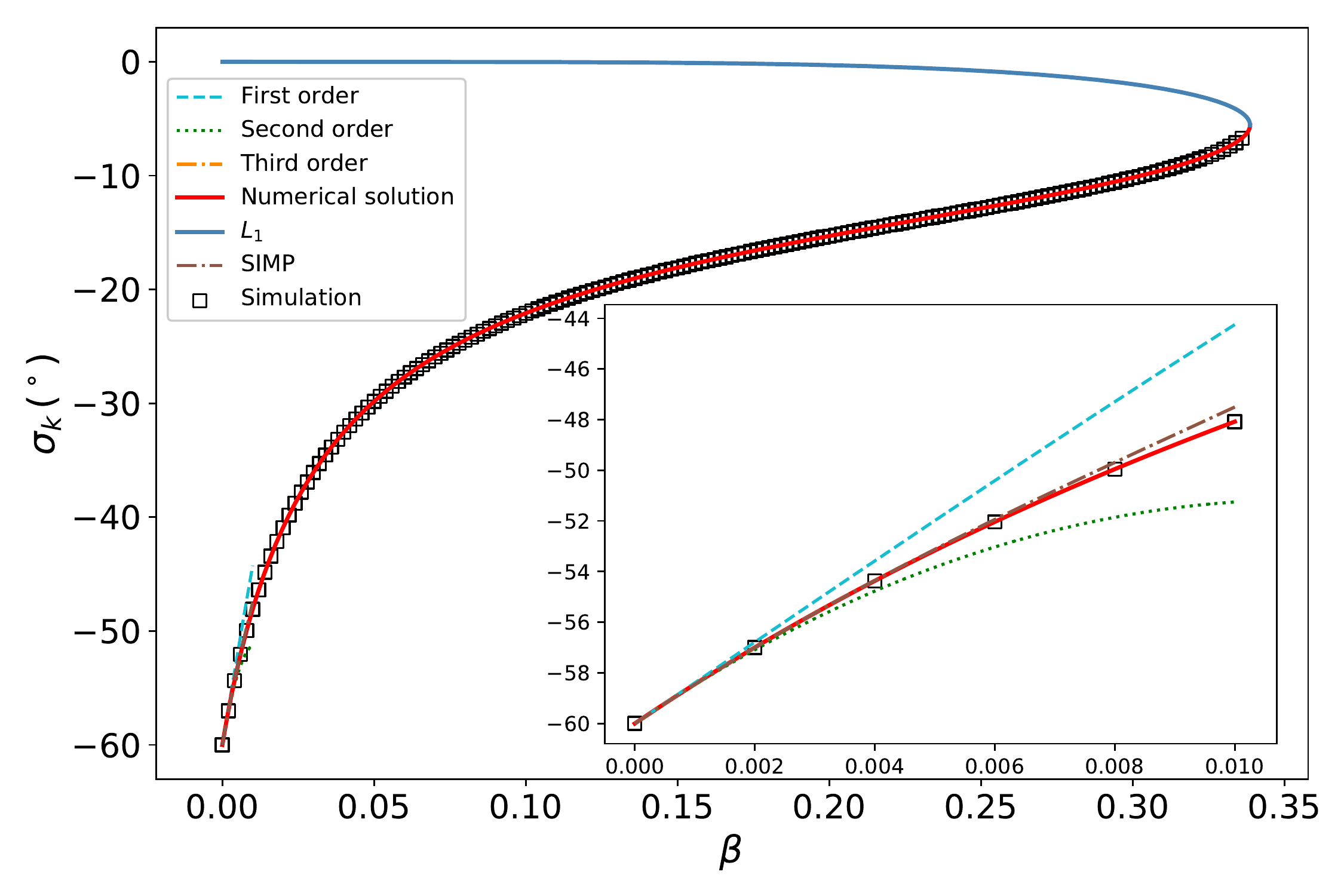}}
\caption{Resonant angle of the libration centers ($\sigma_k$) for $L_4$ (\textit{upper}) and $L_5$
        (\textit{lower}) of Venus from different solutions. The positions of the libration centers for $L_3$ and $L_1$ obtained from the numerical solutions to the equations of motion are also presented in the \textit{upper} and \textit{lower} panels, respectively. `SIMP' represents the results calculated from the simplified approximation formulae Eqs. (\ref{eqn:gensol}) and (\ref{eqn:simp}). The results from approximation formulae for $L_5$ are only shown for $0<\beta<0.01$ (we insert a window to zoom in) because their deviations from the numerical solutions are too large for $\beta>0.01$.}
\label{fig:libcen}
\end{figure}

Equations (\ref{eqn:x00}) and (\ref{eqn:y00}) show that the locations of $L_4$ and $L_5$ are symmetric for $\gamma_c=0$. The distance from both $L_4$ and $L_5$ to the Sun is between 0 and 1 for $0<\beta<1$. As the effect of solar radiation pressure, the triangular Lagrangian points approach the Sun for increasing values of $\beta$, while the distance from the planet remains unchanged. This causes a larger resonant angle for $L_4$ ($>60^\circ$) and a smaller resonant angle for $L_5$ ($<300^\circ$). However, when we take the P-R-S effect into consideration, Eq. (\ref{eqn:r2}) tells us that $L_4$ recedes from the planet while $L_5$ approaches ($y>0$ for $L_4$ while $y<0$ for $L_5$). This leads to the asymmetry between $L_4$ and $L_5$. Therefore, the combined effect of solar radiation pressure and the P-R-S effect always drives $L_4$ away from the planet. At the same time, we always have $r_1<\delta$ from (\ref{eqn:r12}) for $L_4$. This causes the resonant angle of $L_4$ to increase rapidly with $\beta$. The situation is much more complicated for $L_5$. To the first order (see (\ref{eqn:fodsol1})), $r_1$ is always larger than $\delta$, however, which causes the resonant angle to increase. In this case, solar radiation pressure competes with the P-R-S effect to slow down the increase or decrease of the resonant angle, which depends on the value of parameter $\beta$ and $c_v$. We recall Eq. (\ref{eqn:cv}), and we find that $c_v$ is inversely proportional to $v_p/c$, where $v_p$ is the velocity of the planet and is close to the velocity of particles $v$. Therefore a larger $c_v$ corresponds to a smaller $v/c$ and thus a weaker P-R-S effect (see Eq. (\ref{eqn:fsp})). When $c_v$ is large, solar radiation pressure is more likely to dominate so that the resonant angle of $L_5$ still decreases with $\beta$. For small enough $c_v$ , however, the P-R-S effect could counteract the effect of solar radiation pressure or even dominate the motion, leading to an increase in resonant angle.

\subsubsection{Application to Venus}

We applied our formulae to Venus. For $L_4$ and $L_5$, in addition to the simplified formulae shown in Eq. (\ref{eqn:simp}) and the original formulae (first-order to third-order solutions) shown in Appendix~\ref{apd:sol}, we can also obtain $(x,y)$ for $L_4$ and $L_5$ by either numerically solving Eqs. (\ref{eqn:r12}) and (\ref{eqn:r2}) via Newton's method (denoted by `numerical solution' hereafter) or numerically integrating the equation of motion directly (denoted by `simulation' hereafter). Then the resonant angle $\sigma_k$ is calculated with the formulae describing the transformation from the barycentric (CRTBP) or heliocentric (ERTBP) Cartesian coordinate system to heliocentric polar coordinate system. We implemented the Newton method in a Python code to compute the numerical solutions and adopted the SyMBA integrator for simulations. In numerical simulations, the positions of libration centers $(\sigma_k,a_k)$ are determined by locating the centers of libration island in the phase portraits on the $(\sigma,a)$ plane for different $\beta$, where the libration amplitudes $\Delta\sigma$ ($max[\sigma]-min[\sigma]$) are minimum. Concretely, we sampled hundreds of orbits with initial conditions on the $(\sigma_0,a_0)$ plane for each $\beta$ and then integrated them for thousands of years. The other orbital elements were set to be the same as those of Venus. The intervals of $\beta$ were set to be $10^{-4}$ and 0.002 for $L_4$ and $L_5$, respectively. Then we delimited the region of libration centers by comparing the libration amplitudes of these orbits. Afterward, the region enclosing the orbits with minimum libration amplitudes was located. Iteratively, we were able to sample and integrate some orbits in this reduced region and repeated the above steps until we reached a desired resolution. Because there is almost no difference in value between $a_k$ and $a_{1:1}$, we always used $a_{1:1}$ to evaluate $a_k$  for simplicity.

The resonant angles for different Lagrangian points $\sigma_k$ from various solutions are shown in Fig.~\ref{fig:libcen}. Apparently, $\sigma_4$ and $\sigma_5$ always increase with $\beta$. The numerical solutions to Eqs. (\ref{eqn:r12}) and (\ref{eqn:r2}) agree perfectly with the result of numerical simulations based on the ERTBP. Both of them show that there are no solutions for $\beta>0.01135$ for $L_4$ and $\beta>0.33865$ for $L_5$ because they have collided with $L_3$ and $L_1$, respectively (see below). However, even the third-order approximation formula cannot determine the real libration center for large $\beta$, especially for $L_5$. Intuitively, this is because the parameter $c_v\,(\approx 8561)$ in this case is not large enough, leading to a relatively large $\gamma_c$ and slow convergence of the expansion. In other words, the high orbital velocity of Venus results in a relatively large $v/c$, which corresponds to a relatively small $c_v$ according to Eq. (\ref{eqn:cv}). This means that it is not very effective to treat the perturbation from the P-R-S effect as an expansion in $\gamma_c$ for large $\beta$. At the same time, small $c_v$ could also cause an increase of the resonant angle of $L_5$ because the P-R-S effect is more effective here, as we mentioned before.

Given corresponding initial positions $(x,y)$, we can also numerically solve Eqs. (\ref{eqn:r12}) and (\ref{eqn:r2}) via Newton's method and then calculate the resonant angle for $L_1$, $L_2$ , and $L_3$. As shown in Fig.~\ref{fig:libcen}, $\sigma_1$ and $\sigma_3$ decrease with $\beta$. Finally, $L_3$ coincides with $L_4$ at $\beta=0.01135$ with $\sigma_3=\sigma_4=108.4^\circ$. Similarly, $L_1$ coincides with $L_5$ at $\beta=0.33865$ with $\sigma_1=\sigma_5=-5.57^\circ$. For larger $\beta$, these Lagrangian points disappear and the corresponding equilibrium solutions no longer exist. $L_2$ is always located in the forth quadrant for any nonzero $\beta$ and $\sigma_2$ varies in an extremely small range ($<5\times10^{-5}$ deg for $0<\beta<1$).

\subsection{Linear stability}\label{subsec:linearstab}

Consider a dust particle that is given a small displacement $(X, Y)$ from the equilibrium point $(X_0, Y_0)$. The linearized equations of motion (see Appendix~\ref{apd:linearstab}) have the form
\begin{equation}\label{eqn:dxax}
        \vec{\dot{{\rm X}}}=\vec{{\rm A}}\vec{{\rm X}},
\end{equation}
where \vec{{\rm A}} is the matrix $(a_{ij}),$ with $a_{ij}$ ($i,j=0...3$) indicating the linear coefficients with respect to $X$, $Y$, $\dot{X}$, and $\dot{Y}$ in Eqs. \ref{eqn:lsa}, \ref{eqn:lsb}, and
\begin{equation}
        \vec{{\rm X}}=\left(
  \begin{array}{c}
    X \\
    Y \\ 
    \dot{X} \\
    \dot{Y}
  \end{array}
\right).   
\end{equation}
The characteristic equation of \vec{{\rm A}} can be written as
\begin{equation}\label{eqn:chareqn}
        \lambda^4+a_3\lambda^3+(1+a_2)\lambda^2+a_1\lambda+9\mu'(1-\mu')b(1-a_0)=0,
\end{equation}
where $a_k$ ($k=0...3$) are expressions obtained from $a_{ij}$ and 
\begin{equation}
        b=1-\frac{\delta^2}{4}.
\end{equation}
Neglecting the P-R-S effect, that is, setting $\gamma_c=0$, we have $a_i\,(i=0...3)=0$. Then Eq. (\ref{eqn:chareqn}) becomes
\begin{equation}\label{eqn:chareqns}
        \lambda^4+\lambda^2+9\mu'(1-\mu')b=0
\end{equation}
with the solutions
\begin{equation}\label{eqn:chareqnssol}
        \lambda_0=\pm\mathrm{i}\left\{\frac{1}{2}\pm\left[\frac{1}{4}-9\mu'(1-\mu')b\right]^{1/2}\right\}^{1/2}=\mathrm{i}\,\lambda_r.
\end{equation}
In the absence of nongravitational effects except for solar radiation pressure, $L_4$ and $L_5$ are stable if
\begin{equation}
        36\mu'(1-\mu')b\leq1.
\end{equation} 
This constraint is satisfied for all $b$ in $(0,1)$ when 
\begin{equation}
        \mu'\leq\frac{1}{6} \left(3-2 \sqrt{2}\right)\approx0.0286.
\end{equation}

In the presence of the P-R-S effect, we can also solve Eq. (\ref{eqn:chareqn}) by expanding the parameters $a_k$ in $\gamma_c$. The second-order solution is presented in Appendix~\ref{apd:linearstab}. We note that there always exists a positive real part for $\lambda$ when $\gamma_c>0$, which means that $L_4$ and $L_5$ are unstable when we consider the P-R-S effect.

Taking $\mu'\approx0$, the e-folding time of dust particles, which is the time interval in which the oscillation around the Lagrangian points increases by a factor of $e$, can be given by
\begin{equation}\label{eqn:eft}
        \frac{1}{T}\equiv{\rm Re}(\lambda)=\frac{3
   \gamma_c (1+s_w) }{\delta ^2}-\frac{7 {\gamma_c}^2(1+s_w)^2 \left(14-5 \delta ^2\right)}{2y_0\,\delta ^2}.
\end{equation}

The first-order term in $\gamma_c$ has been derived in \citetads{1980ApJ...238..337S}, and we developed it to the second order. The second-order term is opposite for $L_4$ and $L_5$ because they have opposite $y_0$. Therefore the e-folding time for $L_4$ is slightly longer than that for $L_5$. We also extended the solution to the third order. However, $\mu'=0$ is a singularity for the third-order solution, so that we cannot give a similar simplified formula for the e-folding time. The linear stability analysis suggests that the stability of triangular Lagrangian points, which is reflected by the e-folding time, declines with $\beta$. This qualitative result can be confirmed by numerical simulations. Quantitatively, however, the e-folding time is not necessarily equivalent to the lifespan of particles, which is the time duration of staying in the 1:1 MMR region because the linear stability analysis is conducted in a small neighborhood around the equilibrium point. For Venus, the lifespans are much longer than the e-folding time (see Sect.~\ref{subsec:ns}).

\begin{figure}
        \centering
    \resizebox{\hsize}{!}{\includegraphics{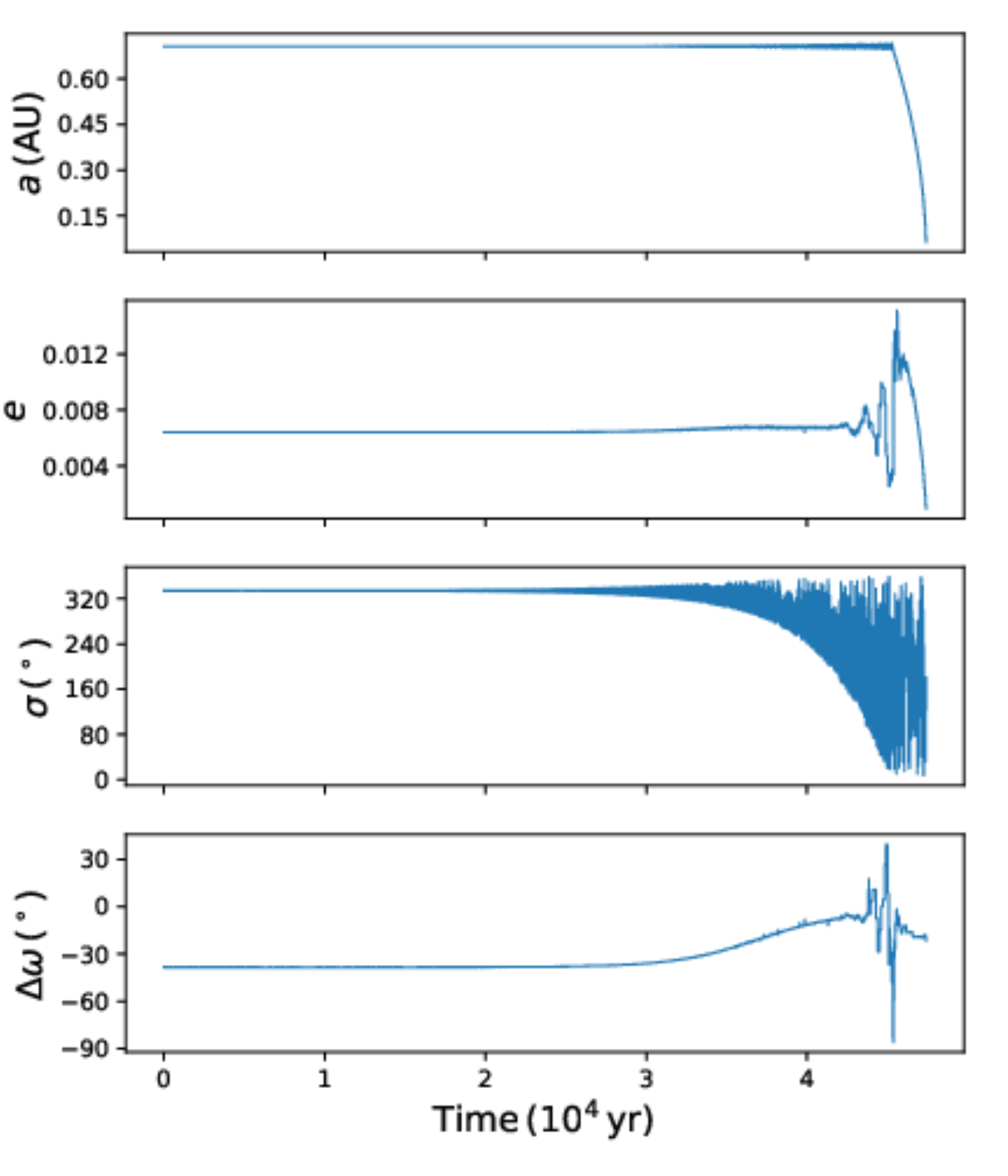}}
                \caption{Orbital evolution of particles in $L_5$ with $\beta=0.07$ ($R=2.9\,{\rm \mu m}$) based on the ERTBP. The orbit is initially placed in the libration center (see more details in the text).}
        \label{fig:exmporb1}
        \end{figure}
        
In Fig.~\ref{fig:exmporb1} we present an example describing the orbital evolution of a particle in $L_5$ with $\beta=0.07$. The initial conditions were set to be in the libration center $(a_k,e_k,i_{\rm V},\Delta\omega_k,\Omega_{\rm V},\sigma_k)=(0.706\,{\rm AU}, 0.0064, 3.39^\circ, -38.45^\circ, 76.68^\circ, -25.93^\circ)$ in the ERTBP (see Sec.~\ref{subsec:ns}). The orbit librated around the libration center with extremely small amplitudes at the beginning. After about 30 kyr, the variations in semimajor axis and resonant angle started to increase. Then the eccentricity changed dramatically, and finally, the orbit escaped at about 45 kyr. As mentioned in \citetads{2019CeMDA.131...49L}, the inclination and longitude of ascending node stay more or less unchanged in the uncharged problem before the orbits escape from the co-orbital region. According to Eq. (\ref{eqn:eft}), the e-folding time is about 3 kyr for $\beta=0.07$ for $L_5$. It is much shorter than the lifespan (45 kyr), as we mentioned before. 

\subsection{Numerical study. ERTBP and N-body}\label{subsec:ns}

\begin{figure}
        \centering
    \resizebox{\hsize}{!}{\includegraphics{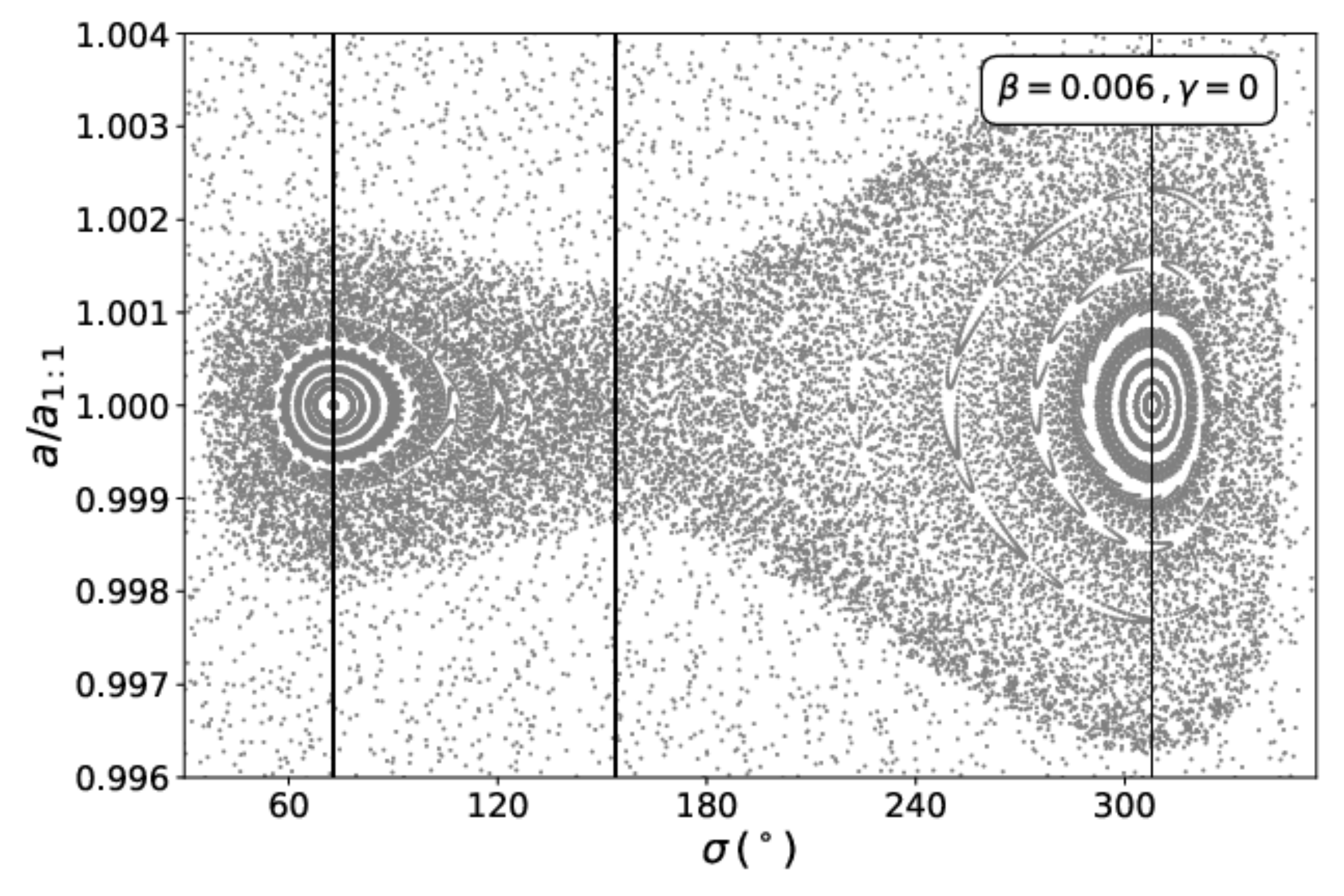}}
    \resizebox{\hsize}{!}{\includegraphics{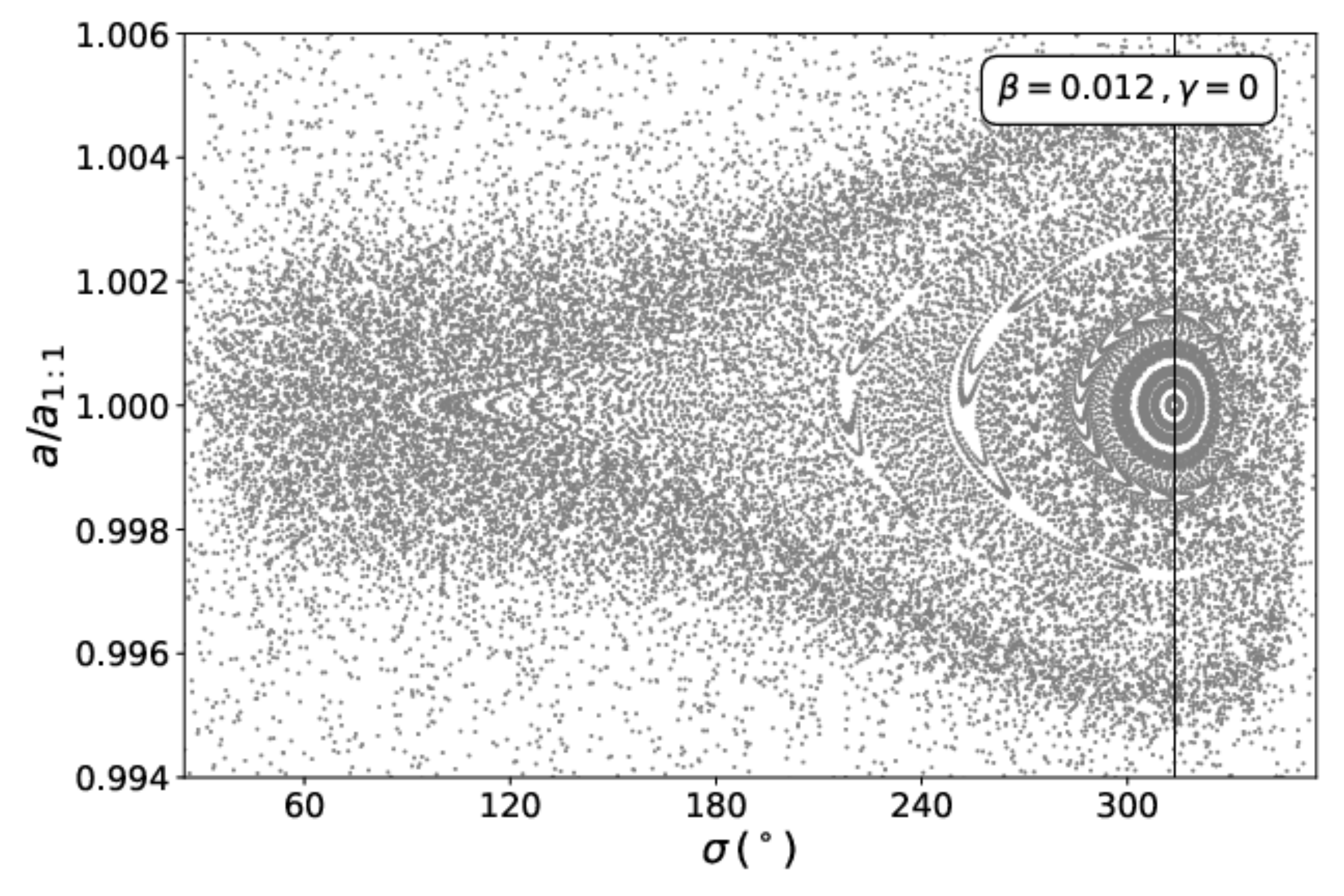}}
                \caption{Phase portrait on the $(\sigma,a)$ plane. The \textit{upper} panel is obtained for $\beta=0.006,$ while the \textit{lower} panel is obtained for $\beta=0.012$. The vertical lines indicate $\sigma_k$ obtained from numerical simulations ($L_4$ and $L_5$) and numerical solutions to the equations of motion ($L_3$), which are $72.8^\circ$ ($L_4$), $153.7^\circ$ ($L_3$) and $308.0^\circ$ ($L_5$) for $\beta=0.006$, $313.6^\circ$ ($L_5$) for $\beta=0.012$.}
                \label{fig:asig}
        \end{figure}

To better illustrate how solar radiation pressure and P-R-S effect affect the dynamical behavior of orbits in 1:1 MMR, we constructed the phase portraits for different values of $\beta$ on the $(\sigma,a)$ plane. Fig.~\ref{fig:asig} presents a comparison between $\beta=0.006$ and $\beta=0.012$. Sixty orbits initially equally spaced $(\Delta{\sigma_0}=6^\circ)$ along $a=a_{1:1}$ were integrated for 10 kyr for each plot within the framework of the ERTBP. The typical phase portrait on the $(\sigma,a)$ plane without any nongravitational effect is shown in Fig.~\ref{fig:phport}, where the locations of $L_4$ and $L_5$ are symmetric and $L_3$ is located at $180^\circ$. However, the presence of $\beta$ brings in an asymmetry to the phase portrait. There exists a displacement of $L_3$, $L_4$ and $L_5$ for $0<\beta<0.01135$, as the example of $\beta=0.006$ shows. The value of $\sigma$ corresponding to $L_3$ decreases to $153.7^\circ$ , while $L_4$ ($72.8^\circ$) and $L_5$ ($308.0^\circ$) shift to the right. These data are obtained from numerical simulations ($L_4$ and $L_5$) and numerical solutions to the equations of motion ($L_3$). Apparently, they perfectly depict the locations of libration points and saddles in the phase portraits (indicated by vertical lines). The perturbation caused by the P-R-S effect gives rise to chaos because the dispersion of orbits in the phase portraits is visible, especially for those far away from triangular Lagrangian points. As mentioned in Sect.~\ref{subsec:loc}, for $\beta>0.01135$, $L_3$ and $L_4$ merge and disappear. Along with this, the tadpole region of $L_4$ vanishes. This is clearly shown by the example of $\beta=0.012$, in which the value of $\sigma$ for $L_5$ is $313.6^\circ$. For all $\beta$, the positions of equilibrium points on the $a$ axis are very close to $a_{1:1}$.  

\begin{figure}
        \centering
    \resizebox{\hsize}{!}{\includegraphics{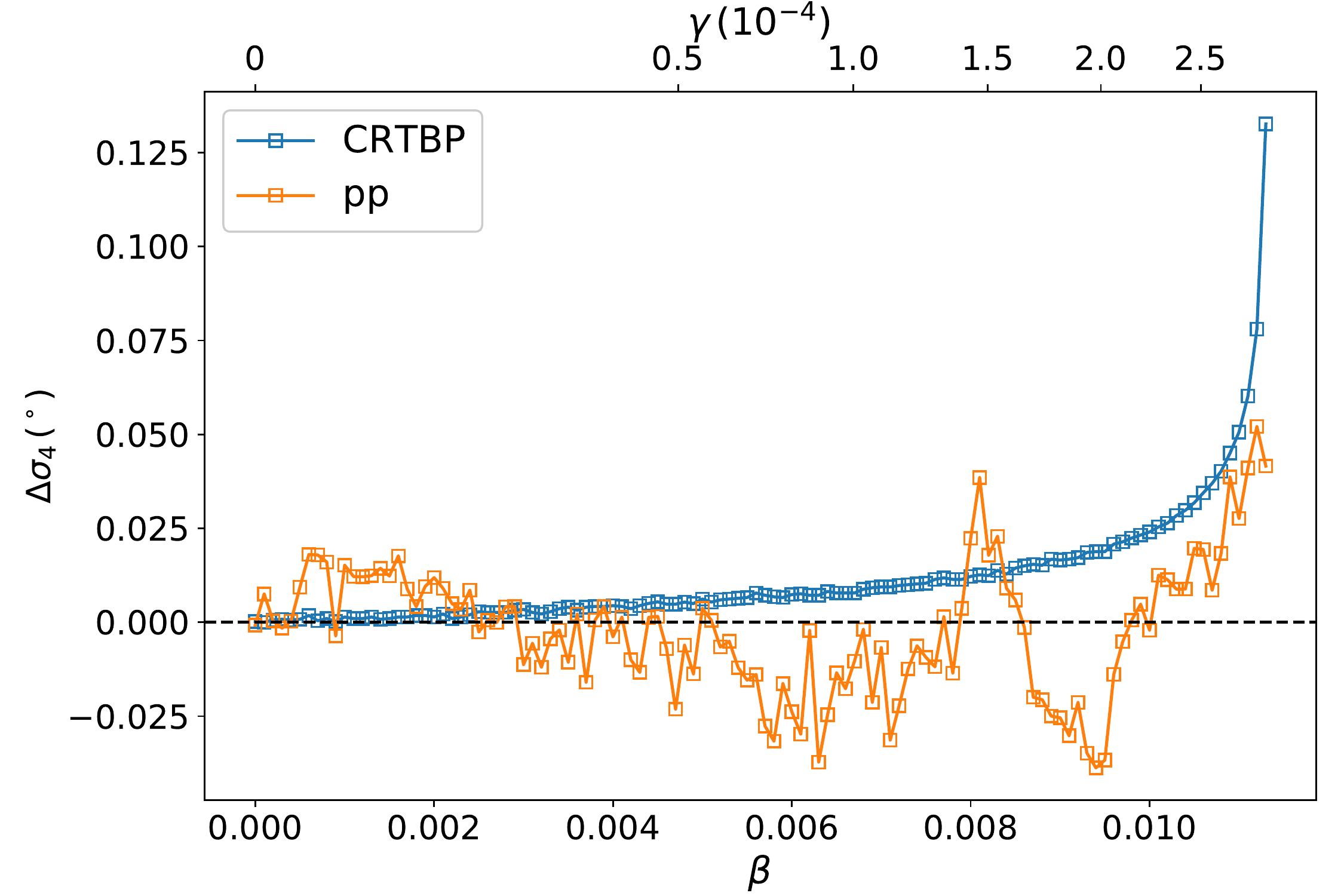}}
    \resizebox{\hsize}{!}{\includegraphics{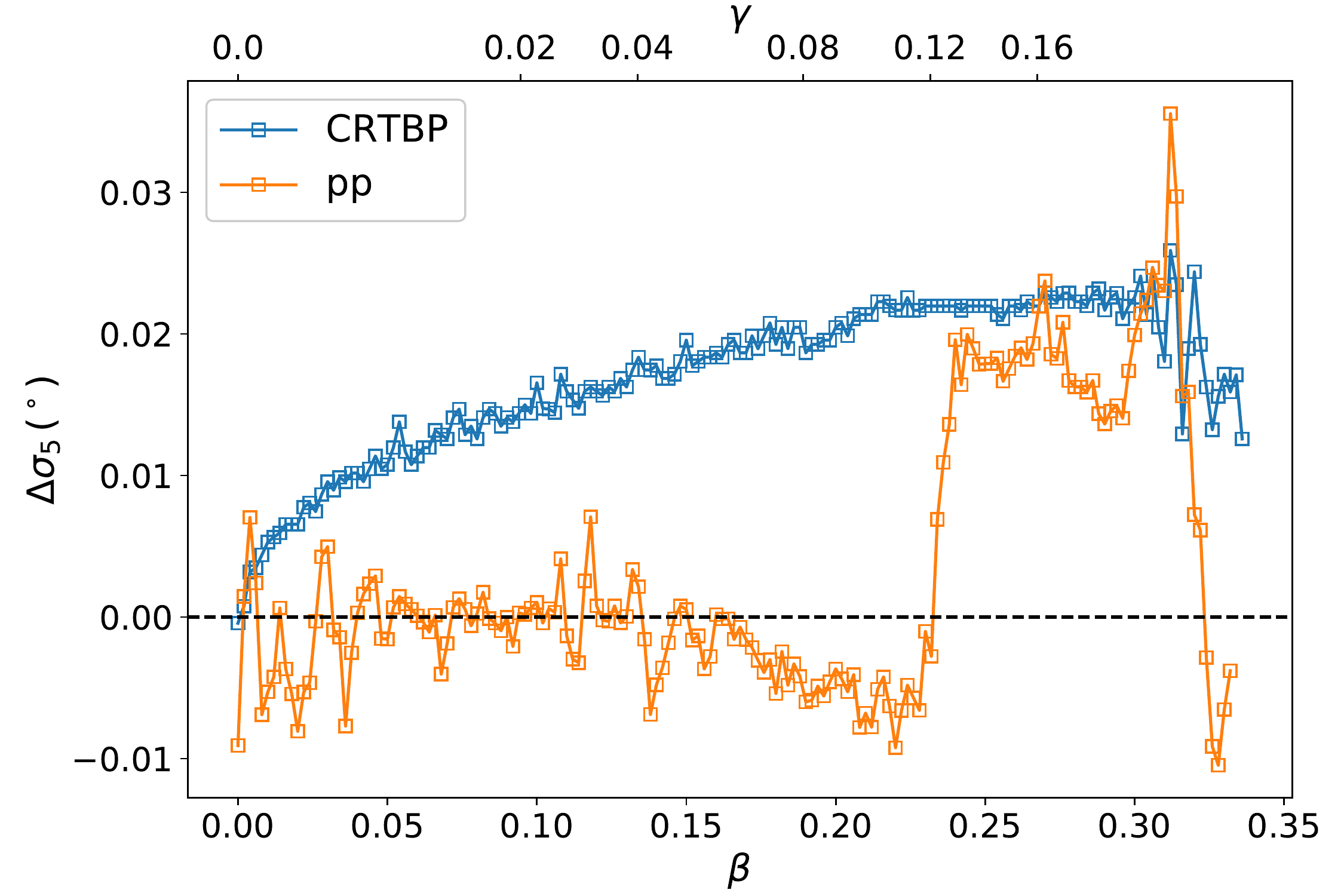}}
                \caption{Difference of resonant angles of libration centers ($\Delta\sigma_k$) between different models and the ERTBP for $L_4$ (\textit{upper}) and $L_5$ (\textit{lower}). `pp' means the planetary perturbations.}
                \label{fig:deltasiga}
        \end{figure}

In Sect.~\ref{subsec:loc} we have also shown a good agreement between the results from the analysis of the CRTBP and numerical simulations in the ERTBP (see Fig.~\ref{fig:libcen}) only with a slight difference between them (denoted by $\Delta\sigma_k$) that is due to the eccentricity of Venus. We calculated the difference in $\sigma_k$ between the numerical solutions (in the CRTBP) and simulations (in the ERTBP). Fig.~\ref{fig:deltasiga} shows that $\sigma_k$ in the CRTBP is always larger than that in the ERTBP, and this kind of difference increases with $\beta$ generally. The eccentricity of Venus has a larger effect for $L_4$ (up to $0.13^\circ$) than $L_5$ (up to $0.025^\circ$). We also note smooth trends of $\Delta\sigma_k$ versus $\beta$, especially for small $\beta$. We assume that an expansion similar to that in Sec.~\ref{subsec:loc} could be conducted to solve the equations of motion within the framework of the ERTBP. Some additional term involved with $\beta$ should appear due to the eccentricity of Venus.

We only considered the case of three-body problem so far. For 1:1 MMR with Venus, the perturbation from the other planets in the Solar System could play an important role in the dynamical evolution of dust orbits. In our paper about Venus Trojans (in prep.), we conduct a comprehensive investigation of the resonance mechanism in the complete model that includes all planets of the Solar System. In Fig.~\ref{fig:deltasiga} we also present the deviations in $\sigma_k$ between the complete model and ERTBP. In contrast to the CRTBP, the planetary perturbations giving rise to the onset of chaos in the whole vicinity of the equilibrium points cause disordered deviations to $\sigma_k$. However, the planetary perturbations result in a comparable shift $\Delta\sigma_k$ as the CRTBP.

As revealed by Fig.~\ref{fig:asig}, in addition to the resonant angle of libration centers $\sigma_k$, the libration width of tadpole regions ($\delta{a}$) also varies with $\beta$. Obviously, the tadpole region of $L_4$ decreases with increasing values of $\beta$ until it disappears completely. In contrast, the tadpole cloud of $L_5$ expands with $\beta$. In practice, we classified the orbits starting in the vicinity of $L_4$ that never exceed $L_3$ during the integration into the tadpole region of $L_4$. Similarly, the tadpole region of $L_5$ consists of the orbits starting in the vicinity of $L_5$ that never go below $L_3$. Many orbits were integrated to precisely delimit the boundary of the tadpole region. In our simulations we used the values of $\sigma$ for $L_3$ obtained from the numerical solutions to the equations of motion (blue curve in the top panel of Fig.~\ref{fig:libcen}) in order to classify the tadpole regions of $L_4$ and $L_5$. As we mentioned in Sec.~\ref{subsec:loc}, $L_3$ and $L_4$ coincide and disappear for $\beta>0.01135$. In this case, we can no longer define the above-mentioned libration width. Finally, the libration width of each triangular Lagrangian point is defined as the maximum difference in semimajor axis of the orbits belonging to the corresponding tadpole region.

\begin{figure}
        \centering
    \resizebox{\hsize}{!}{\includegraphics{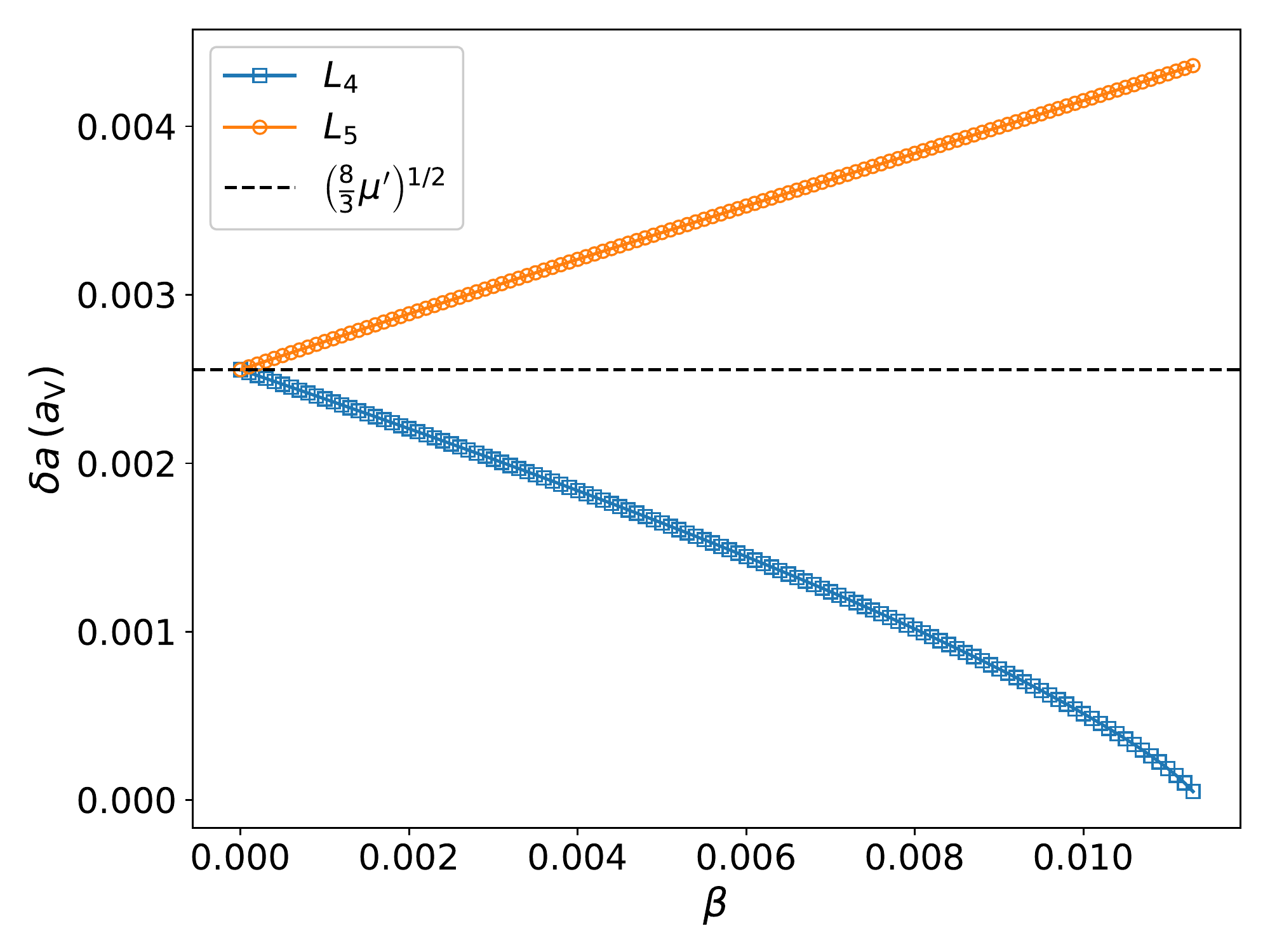}}
                \caption{Libration width of tadpole regions for $L_4$ (blue) and $L_5$ (orange). The dashed line represents the result from the theoretical formula for $\beta=0$. The interval of $\beta$ for $L_5$ is set to be the same as that of $L_4$ in this case ($10^{-4}$).}\label{fig:libwidth}
\end{figure}

According to \citetads{1999ssd..book.....M}, massless particles are thought to move on tadpole orbits if $\delta{r}\leqslant(8\mu'/3)^{1/2}\,a_{\rm V}=2.555\times10^{-3}\,a_{\rm V}$, where $\delta{r}$ is the radial separation of particles from Venus. This condition holds for both $L_4$ and $L_5$ if $\beta=0$, as shown in Fig.~\ref{fig:libwidth}. This figure also shows that the libration width of $L_4$ increases with $\beta,$ while the width of $L_5$ decreases with $\beta$. The corrections are almost linear, especially for $L_5$, where the slope is lower (0.158 v.s. $-0.184$ by linear fitting). The libration width of $L_4$ decreases to zero at $\beta=0.01135$ and the tadpole region around $L_4$ vanishes. At the same time, $\delta{a}$ of $L_5$ increases to $4.360\times10^{-3}\,a_{\rm V}$.

The planetary perturbations could affect the size of tadpole regions by both changing the location of $L_3$ and affecting the dynamical behavior near the boundary of tadpole regions. The difference is unremarkable in most cases, however. The maximum deviation is less than $5\times10^{-7}\,a_{\rm V}$. Considering that the libration width tends to 0 for $L_4$ when $\beta$ tends to 0.01135, however, the planetary perturbations play an increasingly important role with increasing values of $\beta$.

\begin{figure}
        \centering
    \resizebox{\hsize}{!}{\includegraphics{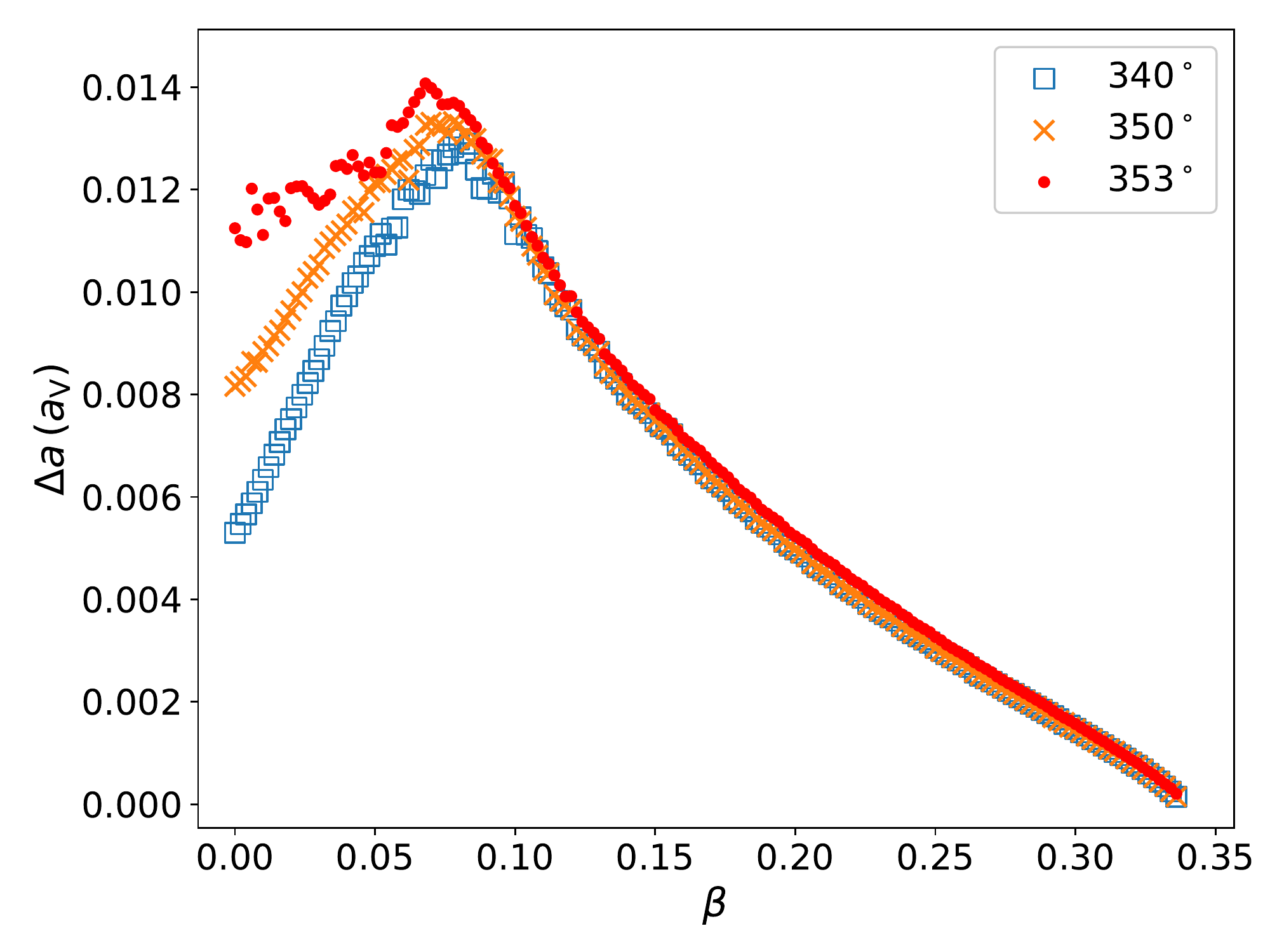}}
                \caption{Libration width of 1:1 MMR with Venus for different thresholds. See text for more details.}
                \label{fig:libwidth11}
        \end{figure}

When we include the horseshoe orbits, determining the extended libration width of the 1:1 MMR region $(\Delta{a})$ becomes more complicated because such orbits are chaotic due to the P-R-S effect. We set different criteria to describe the boundary of the 1:1 MMR region of Venus. Three different thresholds ($340^\circ,\,350^\circ, \text{and}\,353^\circ$) indicating the maximum allowed libration amplitudes were chosen. The results shown in Fig.~\ref{fig:libwidth11} suggest that the libration width of the 1:1 MMR increases with $\beta$ first and then decreases. It is obvious that a higher threshold value of libration amplitudes produces a smaller turning point. In the region before the turning point, the libration width is heavily dependent on the threshold. This is due to the chaotic nature near the boundary, which also causes the roughness of the curve for a high threshold such as $353^\circ$. Meantime, the gradual rise of the width is caused by the expansion of the stable region around $L_5$ (see Fig.~\ref{fig:asig}). After the turning point, the increasing perturbation due to the P-R-S effect gradually destabilizes the orbits in the 1:1 MMR region. Only the particles closely orbiting $L_5$ could survive, so the dependence of the libration width on the threshold is much weaker. 

\begin{figure}
        \centering
    \resizebox{\hsize}{!}{\includegraphics{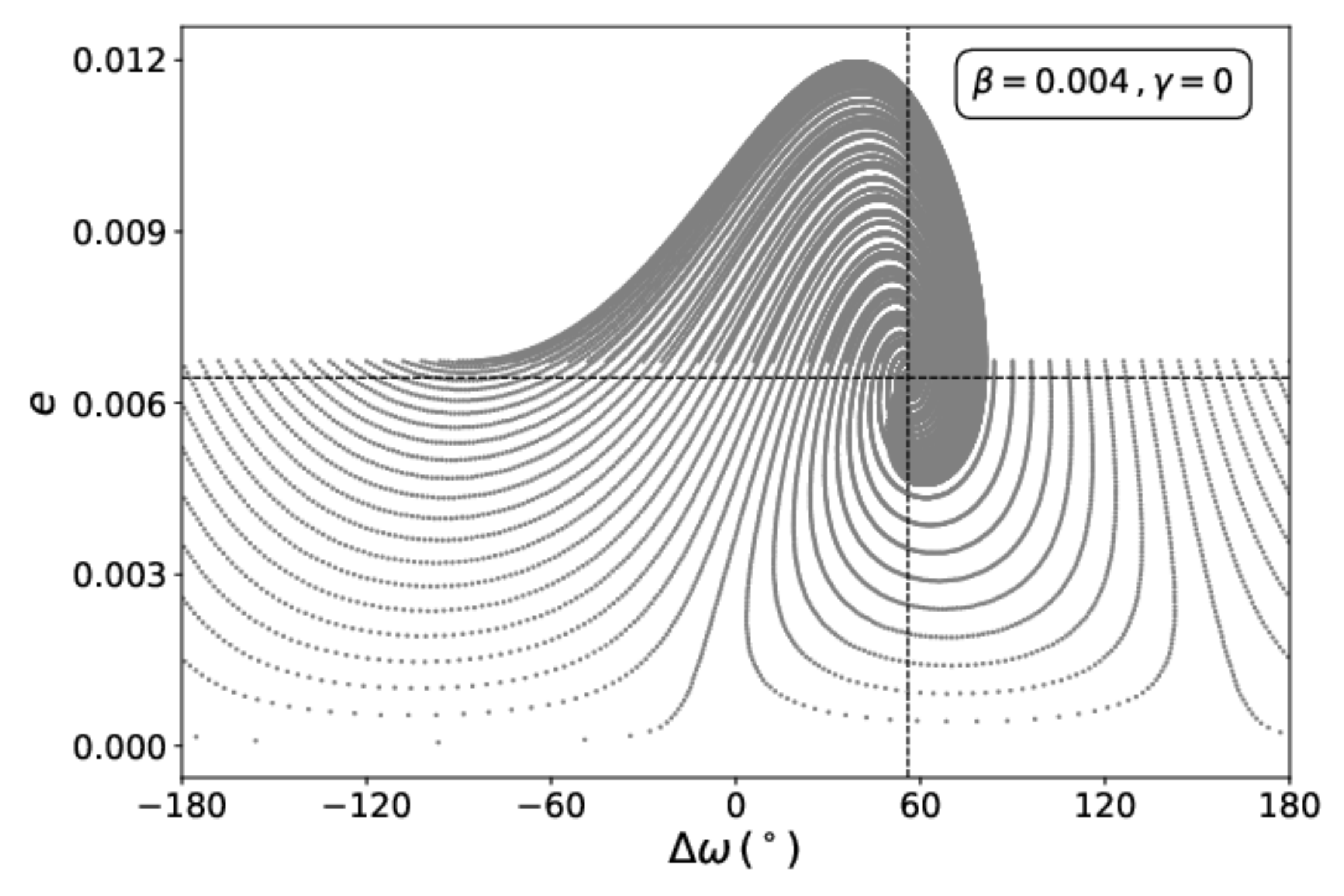}}
    \resizebox{\hsize}{!}{\includegraphics{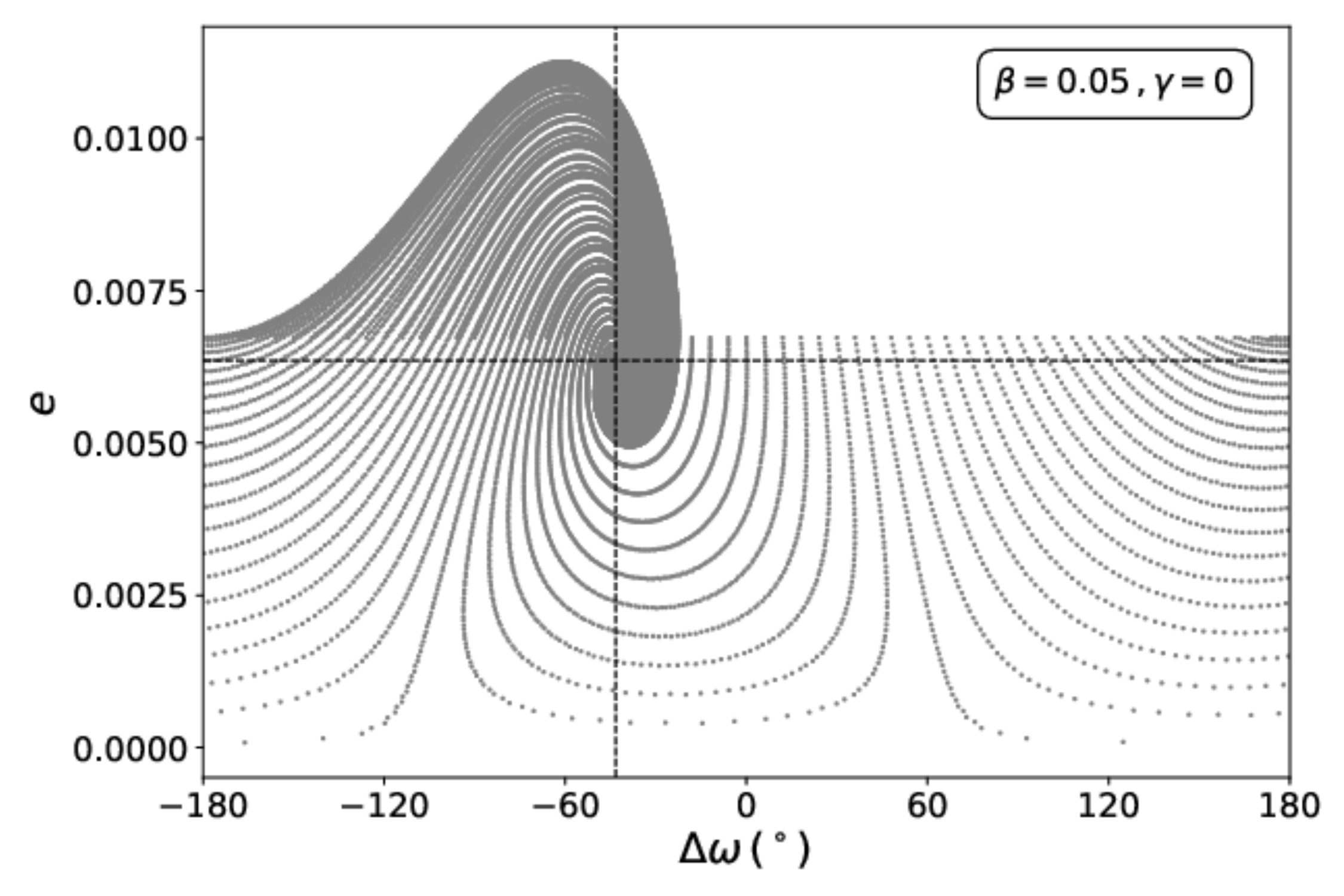}}
                \caption{Phase portrait on the $(\Delta\omega,e)$ plane. The \textit{upper} panel is for $\beta=0.004$ for $L_4$ , and the \textit{lower} panel is for $\beta=0.05$ for $L_5$. The horizontal and vertical lines indicate $e_k$ and $\Delta\omega_k$ obtained by minimizing the libration amplitudes, which are 0.00644 and $55.90^\circ$ for the \textit{upper} panel, 0.006355 and $-43.24^\circ$ for the \textit{lower} panel.}
                \label{fig:edom}
        \end{figure}

After we determined the positions of libration centers $(\sigma_k,a_k),$  we studied the change in the phase plane $(\Delta\omega,e)$  by modifying the value of $\beta$. In Fig.~\ref{fig:edom} we show on the $(\Delta\omega,e)$ plane the evolution of 60 orbits with initial conditions $\sigma=\sigma_k$, $a=a_k$. According to \citetads{2015Icar..250..249L} and \citetads{2019CeMDA.131...49L}, $(i,\Omega)$ of $L_4$ and $L_5$ are always located at $(i_{\rm V},\Omega_{\rm V})$. These two elements remain unchanged during the integration and are not affected by solar radiation pressure and the P-R-S effect. In comparison with the typical phase portrait shown in Fig.~\ref{fig:phport} for $\beta=0$, solar radiation pressure and the P-R-S effect exert additional forces in the orbital plane on dust particles and cause a deviation of the center $(\Delta\omega_k,e_k)$. In addition, the orbits evolve spirally toward the center in the phase space regardless of the initial conditions. In the two cases we show in Fig.~\ref{fig:edom} ($\beta=0.004$ for $L_4$ and $\beta=0.05$ for $L_5$), the apsidal difference between the libration centers and Venus $\Delta\omega_k$ approaches 0 and the eccentricity $e_k$ gets lower than $e_{\rm V}$ in comparison with the case of $\beta=0$, where $\Delta\omega_k=\pm60^\circ$ and $e_k=e_{\rm V}$.

\begin{figure}
        \centering
    \resizebox{\hsize}{!}{\includegraphics{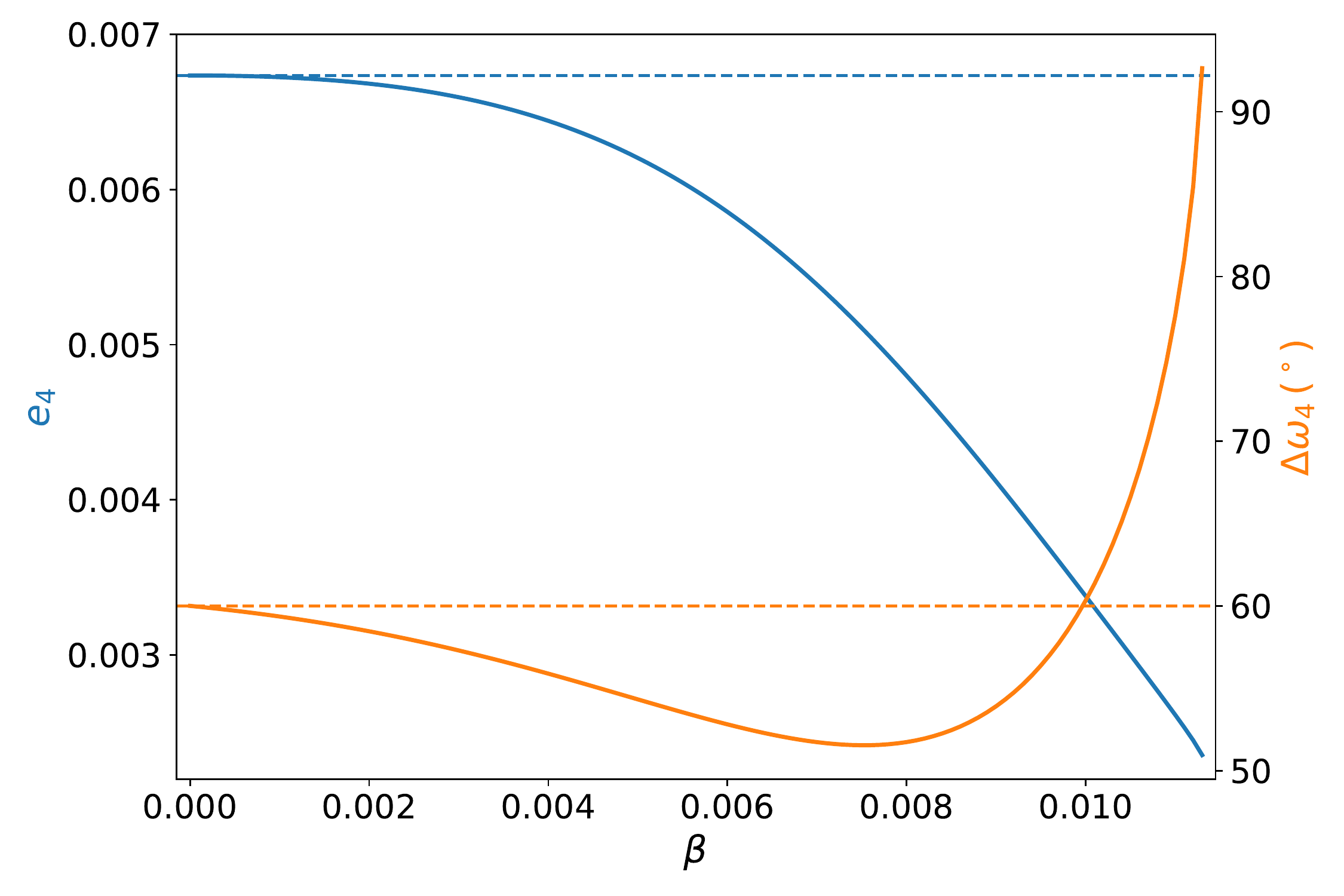}}
    \resizebox{\hsize}{!}{\includegraphics{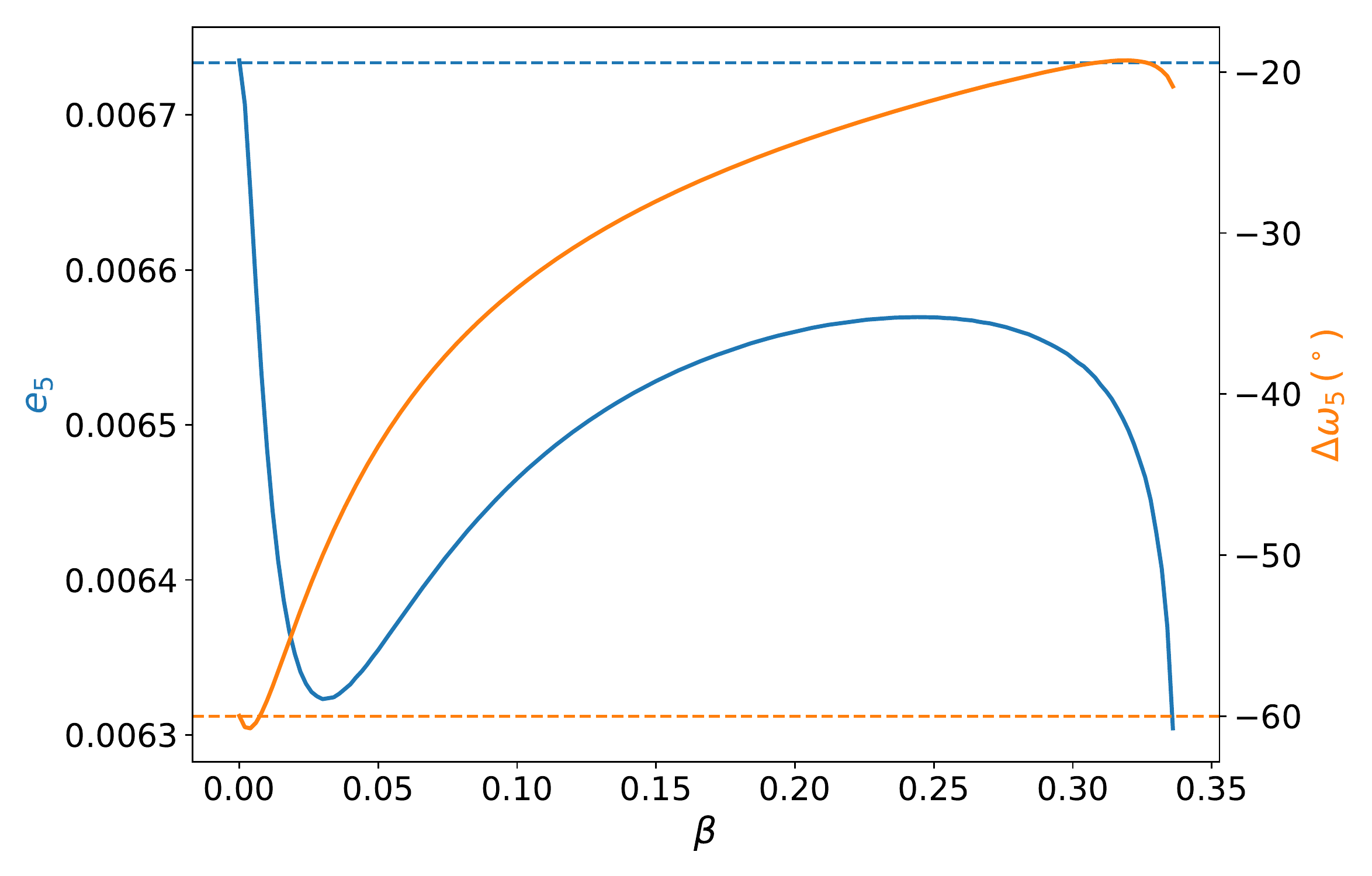}}
                \caption{Eccentricity (blue) and apsidal difference (orange) with the smallest amplitudes for different values of $\beta$ for $L_4$ (\textit{upper}) and $L_5$ (\textit{lower}). The blue horizontal line indicates the eccentricity of Venus (0.00673), and the orange horizontal line indicates $\pm60^\circ$.}
\label{fig:eome}
        \end{figure}

Following the method we used to determine $(\sigma_k,a_k)$ in Sect.~\ref{subsec:loc}, the centers of $(\Delta\omega,e)$ were also precisely determined by minimizing the libration amplitudes in an iterative way. Fig.~\ref{fig:eome} presents the result. For $\beta=0$, the phase portrait is symmetric for $L_4$ and $L_5$ and the center $(\Delta\omega_k,e_k),$ where the amplitudes of $e$ and $\Delta\omega$ are minimum, is located at the point of $(e_{\rm V},\pm60^\circ)$. In the presence of a non-zero $\beta$, we find an asymmetry between $L_4$ and $L_5$. The eccentricity $e_k$ with the smallest amplitudes monotonously decreases with $\beta$ for $L_4$, at a much faster rate than for $L_5$ in the same range of $\beta$ $(0,0.01135)$. Therefore a maximum $\beta$ value of 0.01135 corresponds to a minimum $e_k$ of 0.00235. For dust particles around $L_5$, although $e_k$ could increase with $\beta$ for $0.03<\beta<0.245$, it is still smaller than $e_{\rm V}$ in the whole range to be considered $(0,0.33865)$. In short, the combined effect of solar radiation pressure and the P-R-S effect always tends to reduce the eccentricity of dust particles of any size we considered in the triangular Lagrangian points. Despite a wider range of $\beta$, $e_k$ of dust particles in $L_5$ varies in a narrower range with a minimum value of 0.00630.

 Fig.~\ref{fig:eome} shows that the apsidal difference $\Delta\omega_k$ varies with $\beta$ in a more sophisticated way than $\sigma_k$ does (cf. Fig.~\ref{fig:libcen}). It always decreases fist and then increases with $\beta$ (for $L_5$ it even decreases again for $\beta>0.32$). The (first) turning point is $\beta\approx0.008$ for $L_4$ and $\beta\approx0.005$ for $L_5$. We note that it is possible to observe dust particles of specific sizes moving on orbits with $\Delta\omega_k=\pm 60^\circ$, just like the case where $\beta=0$, but $e_k<e_{\rm V}$. For $L_4$ , this size is 21 ${\rm \mu m}$, corresponding to a $\beta$ of 0.009963, and it is 28 ${\rm \mu m}$ ($\beta=0.00734$) for $L_5$.

\begin{figure}
        \centering
    \resizebox{\hsize}{!}{\includegraphics{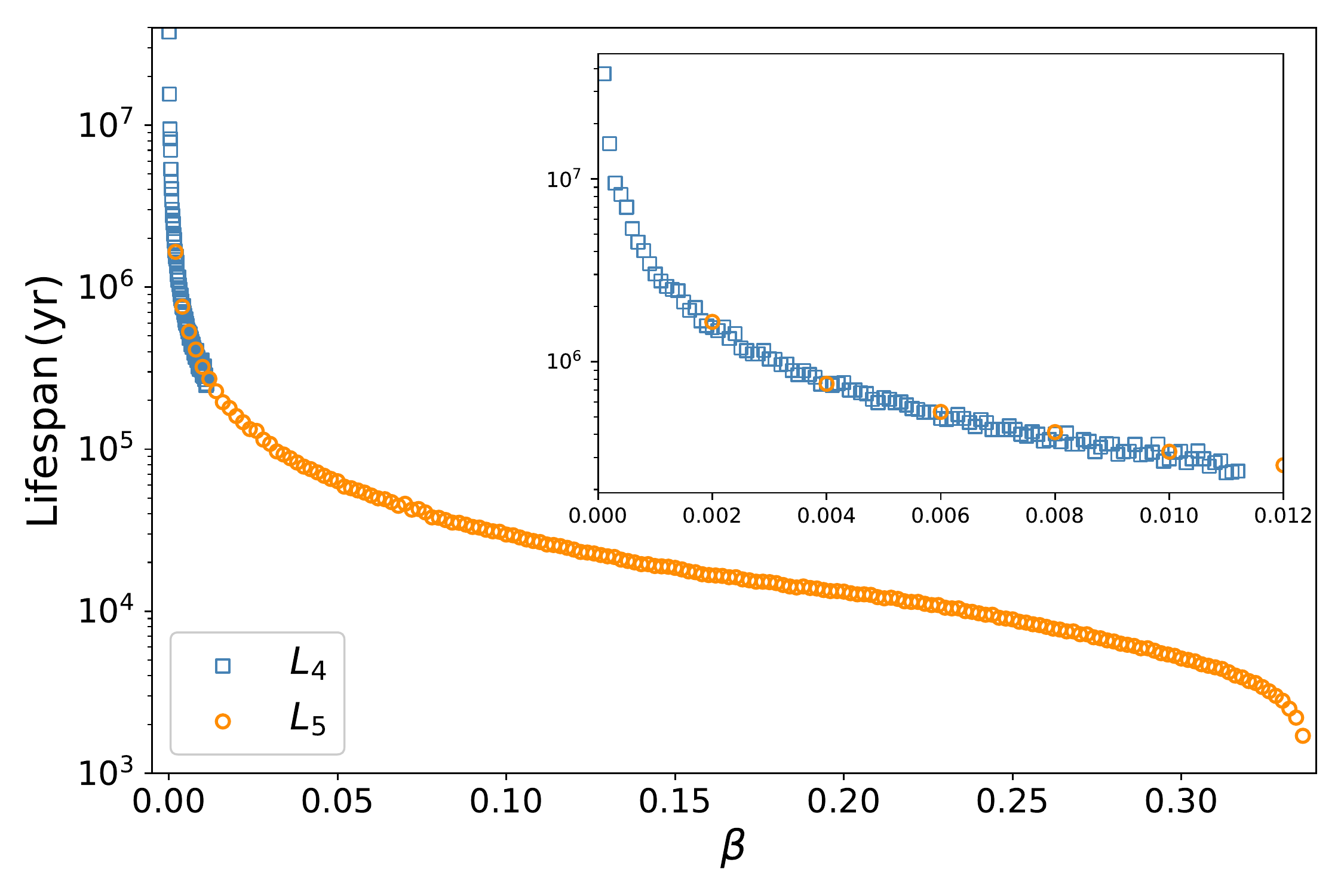}}
                \caption{Lifespans of dust particles in the libration centers of $L_4$ and $L_5$. A window is inserted to zoom in the region of $0<\beta<0.012$.} \label{fig:lf}
        \end{figure}

Although the tadpole region of $L_5$ and the 1:1 MMR region may be enlarged under some conditions (see Fig.~\ref{fig:libwidth} and \ref{fig:libwidth11}), the stability of triangular Lagrangian points is thought to be reduced for increasing $\beta$. To verify this, we integrated the orbits in the libration centers ($a_k,e_k,i_{\rm V},\Delta\omega_k,\Omega_{\rm V} ,\text{and }\sigma_k$) for different $\beta$. The integration of one particle was ended when it escaped the 1:1 MMR region of Venus, and then we recorded the lifespan. After several test simulations, we applied a criterion that if an orbit dissatisfies $|a-a_k|<0.0075\,{\rm AU}$ at any time, it was regarded as escaped. It is feasible to set a threshold of 0.0075 AU, which is slightly higher than the maximum half-width of the 1:1 MMR region (see Fig.~\ref{fig:libwidth11}), because the orbits escaping the 1:1 MMR are scattered away or fall into the star in a short time. Generally, the lifespans shown in Fig.~\ref{fig:lf} decrease with $\beta$, indicating a diminishing stability. The trends of lifespans for $L_4$ and $L_5$ seem to be similar in the range of $\beta<0.01135$. In the CRTBP, the triangular Lagrangian points are dynamically stable for all planets of the Solar System. When we included solar radiation pressure and the P-R-S effect, $L_4$ and $L_5$ were no longer stable. This was shown in the linear stability analysis in Sec.~\ref{subsec:linearstab}. As we also mentioned there, the lifespans here are several times longer than the e-folding time, but these two time scales follow a similar trend with $\beta$. Fig.~\ref{fig:lf} also shows that the lifespan is about 38 Myr for orbits in $L_4$ with $\beta=10^{-4}$ and about 1.8 Myr for orbits in $L_5$ with $\beta=0.002$. As $\beta$ increases, the lifespan drops monotonously in general and decreases to 0.1 Myr when $\beta=0.03$ (only for $L_5$). 

\section{Charged problem}
\label{sec:cp}

In this section we reexamine the results of the previous section including the charge, that is, we focus on the role of the the interplanetary magnetic field on captured charged dust grains in co-orbital resonance with Venus. Given the background magnetic field strength at specific distance, solar rotation rate and solar wind speed (see Table.~\ref{tab:param}), the Lorentz force shown in Eq. (\ref{eqn:fl}) is determined by the charge-to-mass ratio, position, velocity, and the deviation of the magnetic and orbital angular momentum axis.

\subsection{Shift of resonance and capture time}

\begin{figure}
        \centering
    \resizebox{\hsize}{!}{\includegraphics{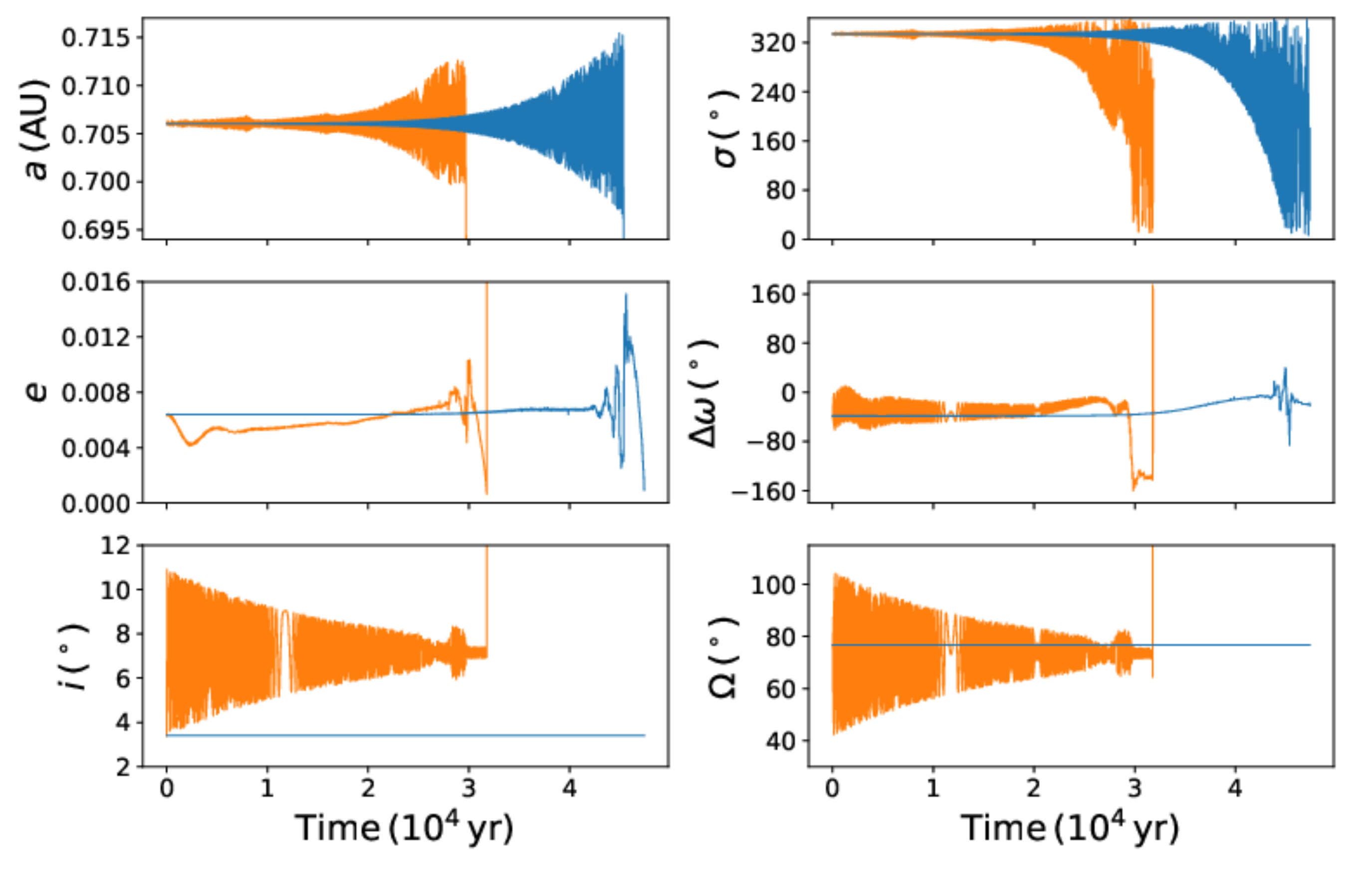}}
                \caption{Orbital evolution of particles initially located in $L_5$ for $\beta=0.07$. Orange curves indicate the model including the Lorentz force with $\gamma=0.011$ (surface potential = 10 V), and the blue curves indicate the same orbit as in Fig.~\ref{fig:exmporb1}, where $\gamma=0$. Both orbits are initially placed in the libration center of the uncharged problem.}
        \label{fig:exmporb2}
        \end{figure}

In consideration of the important effect of the interplanetary magnetic field, especially for small particles \citepads[see e.g.][]{2014PhR...536....1M,2019CeMDA.131...49L}, we introduced the Lorentz force in our simulations. The charge-to-mass ratio $q/m$ (or $\gamma$) is related to $\beta$ through Eq. (\ref{eqn:bqm}), which can be written as $\gamma\approx0.224\,U\beta^2$ assuming $\rho=2.8\,{\rm g/cm^3}$. We adopted the value of the maximum surface potential ($U=10\,{\rm V}$) used in \citetads{2014PhR...536....1M} and \citetads{2019CeMDA.131...49L} in this section, and thus we have $\gamma\approx2.24\,\beta^2$. 

A comparison of the orbital evolution between charged and uncharged particles is shown in Fig.~\ref{fig:exmporb2}. We switched on the Lorentz force for the orbit shown in Fig.~\ref{fig:exmporb1}. The corresponding value of $\gamma$ is 0.011 for $\beta=0.07$. We observe a shorter lifespan for the charged particle ($\sim 30$ kyr). Except for this, however, the semimajor axis and resonant angle have a similar evolution for both particles, although the Lorentz force causes larger libration amplitudes at the very beginning. A close inspection also suggests that under the effect of the Lorentz force, deviations in the locations of libration centers in the phase space exist, especially in $e$ and $\Delta\omega$. Furthermore, $i$ and $\Omega$ are no longer constant during the integration because of the additional normal force exerted by the interplanetary magnetic field. The inclination can be pumped up to $\sim 10^\circ$. The longitude of the ascending node follows a similar evolution as the inclination, librating with decreasing amplitudes.  $i$ and $\Omega$ are coupled, and their variation is confirmed to be related to the deviation of the magnetic axis from the angular momentum axis of the system, the initial inclination, and the ratio $\Omega_s/n$ \citepads{2016ApJ...828...10L,2019CeMDA.131...49L}.

\begin{figure}
        \centering
    \resizebox{\hsize}{!}{\includegraphics{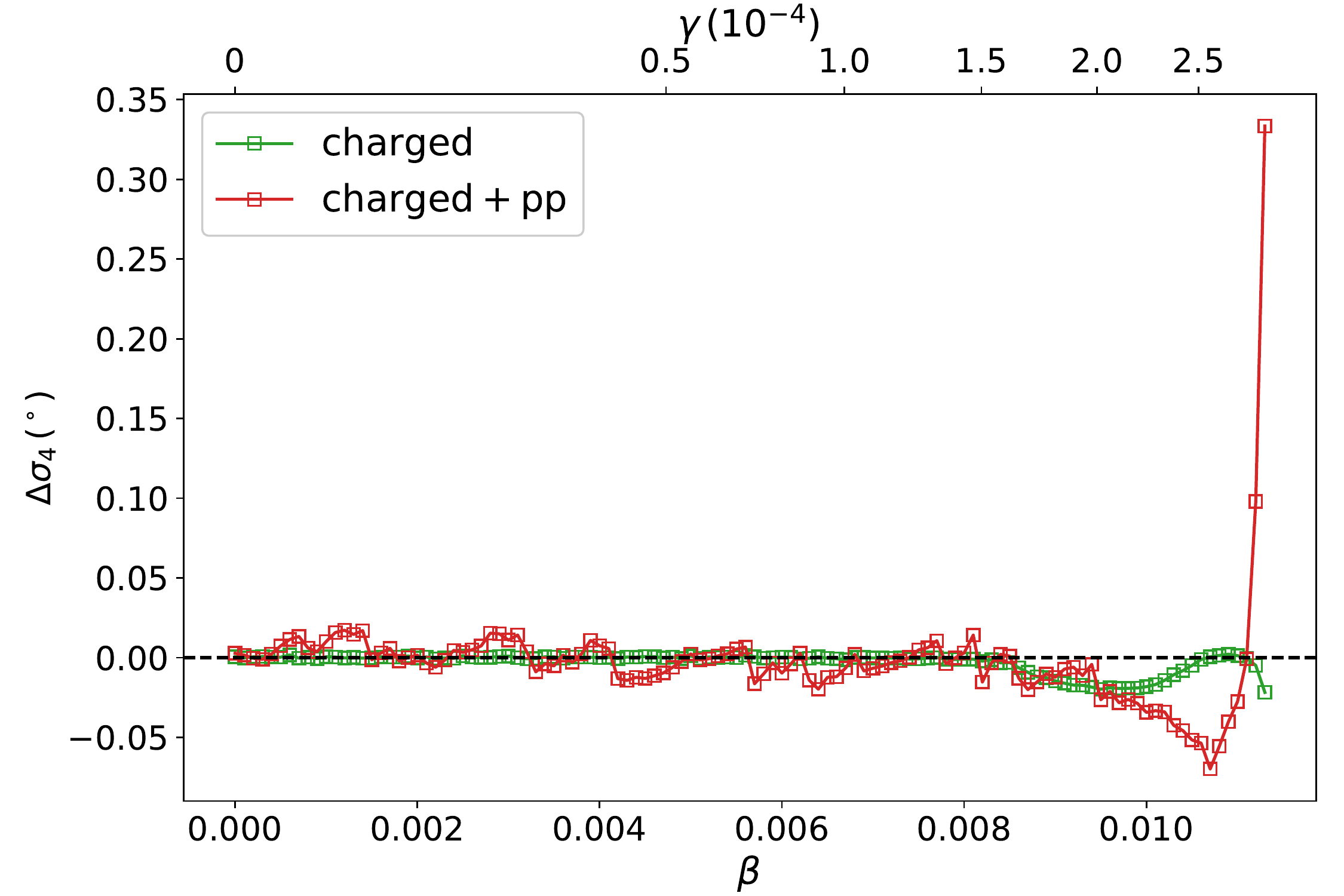}}
    \resizebox{\hsize}{!}{\includegraphics{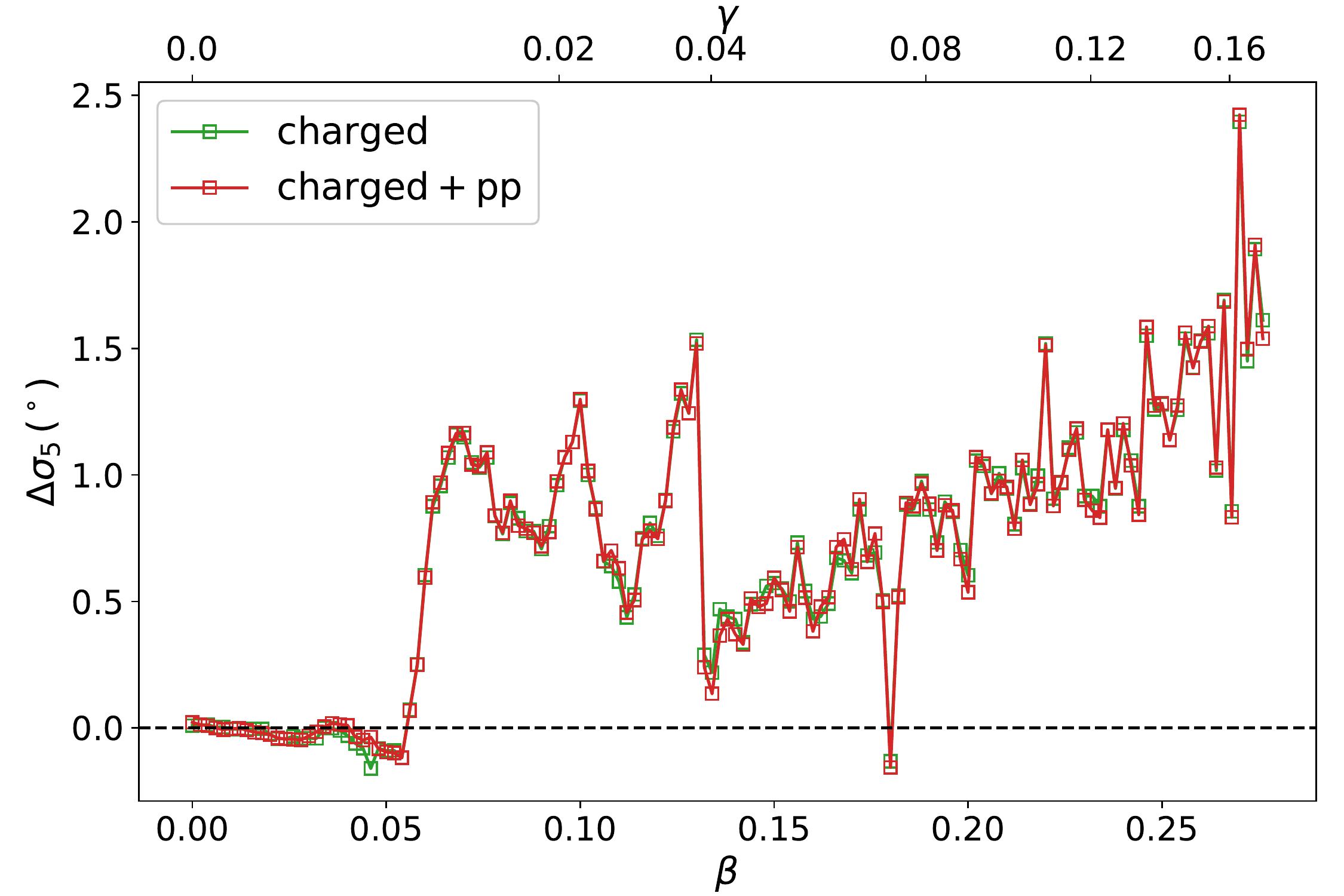}}
                \caption{Differences of resonant angles of libration centers $\Delta\sigma_k$ between different models and the ERTBP for $L_4$ (\textit{upper}) and $L_5$ (\textit{lower}). `Charged' refers to the model including the Lorentz force.}
                \label{fig:deltasigb}
        \end{figure}

The resonant angle of libration centers $\sigma_k$ depends on the Lorentz force. Because we fixed the surface potential, the Lorentz force is larger for smaller particles, whose $\beta$ are also larger, according to $\gamma\approx2.24\,\beta^2$. We repeated the iterative method in Sec.~\ref{subsec:loc} to locate $\sigma_k$ of libration centers. The upper panel of Fig.~\ref{fig:deltasigb} shows that for $L_4$ the Lorentz force has a negligible effect on $\sigma_k$ for $\beta<0.0085$, corresponding to $R>24\,{\rm \mu m}$ and $\gamma<1.6\times10^{-4}$. Even a higher value of $\gamma$ can only cause a deviation no greater than $0.02^\circ$ in $\sigma_4$. This is principally due to the low limit of $\beta$ ($\leq0.01135$) required by the existence of $L_4$ (see Sect.~\ref{subsec:loc}), which corresponds to a small limit of $\gamma$ ($\leq2.9\times10^{-4}$) and thus a small Lorentz force. However, when a higher value of $\beta$  is allowed, as in the case of $L_5$ where $\beta\le0.33865$, the Lorentz force could become larger and affect the position of $L_5$ in a more effective way (see the lower panel of Fig.~\ref{fig:deltasigb}). The maximum difference $\Delta\sigma_k$ could reach $2.4^\circ$, which corresponds to a distance of $4.5\times10^6$ km from the location of the exact libration point in uncharged problem. In addition, for $\beta>0.276,$ no stable points can be found in the phase portrait because of the chaos that is induced by the interplanetary magnetic field, which also causes the chaotic variation in $\Delta\sigma_k$ with $\gamma$.  Fig.~\ref{fig:deltasigb} also shows that in general, $\sigma_k$ in the charged problem seems to be smaller than that in the uncharged problem for $L_4$ (negative $\Delta\sigma_4$) and larger than that in the uncharged problem for $L_5$ (positive $\Delta\sigma_5$) for most $\beta$.

Planetary perturbations might clearly enhance the chaotic behavior in the whole system. The isolated effect of the Lorentz force on the position of $L_4$ (see Fig.~\ref{fig:deltasigb}) is slightly smaller than that of planetary perturbations (see also Fig.~\ref{fig:deltasiga}). However, for $L_5$, of which the $\beta$ range is much wider, the deviation coming from planetary perturbations is negligible in comparison of that from the Lorentz force for $\beta>0.054$ (see the lower panel of Fig.~\ref{fig:deltasigb}). We also took into account the planetary perturbations and Lorentz force in our simulations. When these two forces are comparable in magnitude, the combined effect may produce a larger deviation in $\sigma_k$, just like in the case of $L_4$ (see the top panel in Fig.~\ref{fig:deltasigb}). However, when one of these two forces dominates, the combined effect follows the dominant effect, just like in the case of $L_5$ (see the lower panel in Fig.~\ref{fig:deltasigb}). 

The Lorentz force could further enlarge the tadpole region of $L_5$ while reducing the tadpole region of $L_4$ (not shown here). In comparison with a maximum deviation of $3.5\times10^{-7}\,a_{\rm V}$ in the libration width of $L_5$, the Lorentz force has a much stronger effect on the tadpole region in $L_4$, which can be reduced by up to $8\times10^{-6}\,a_{\rm V}$. Although the deviations are very small, this reduction could be really remarkable (up to 15\%) because the libration width of $L_4$ for uncharged
particles is also very small for large $\beta$.

\begin{figure}
        \centering
    \resizebox{\hsize}{!}{\includegraphics{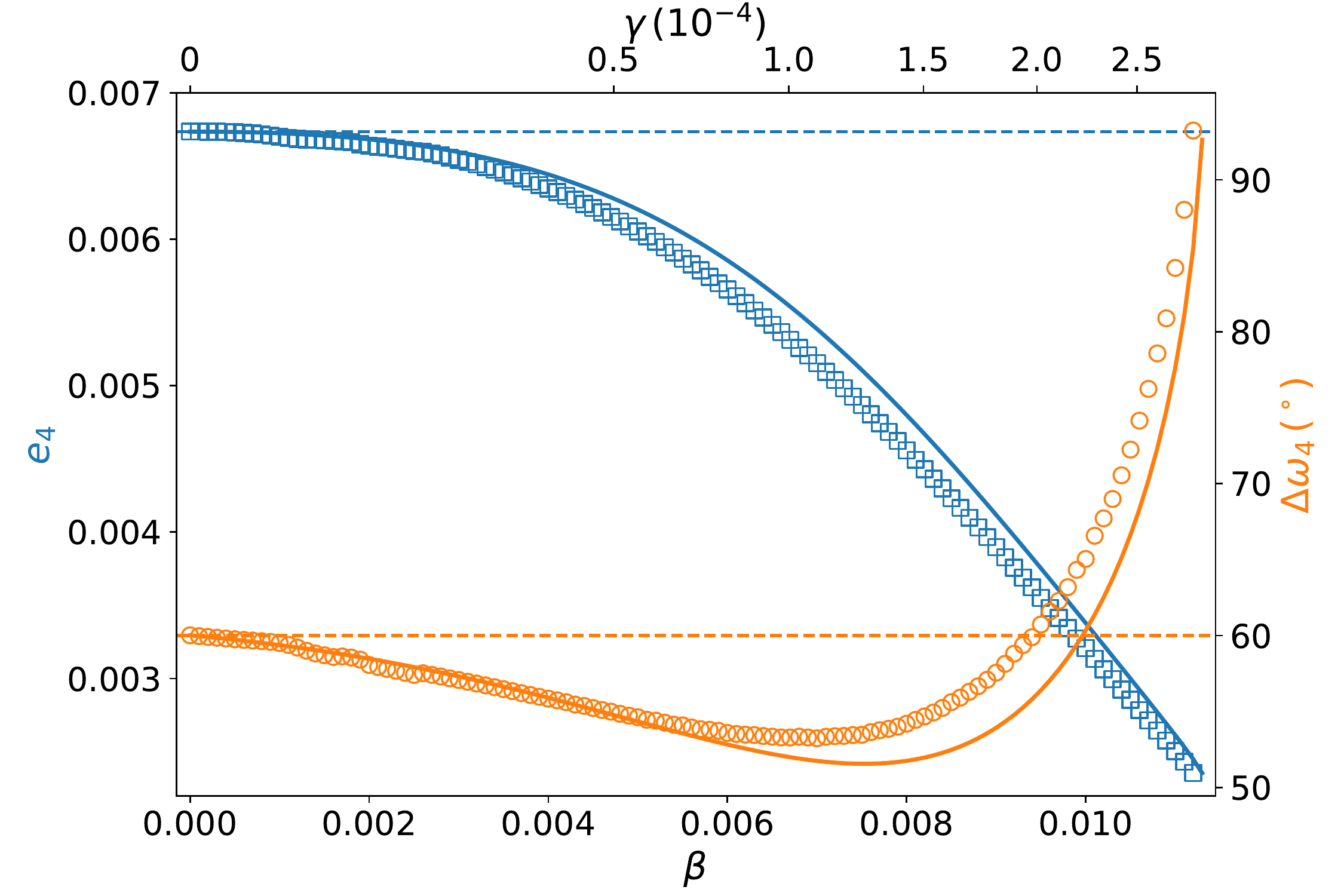}}
    \resizebox{\hsize}{!}{\includegraphics{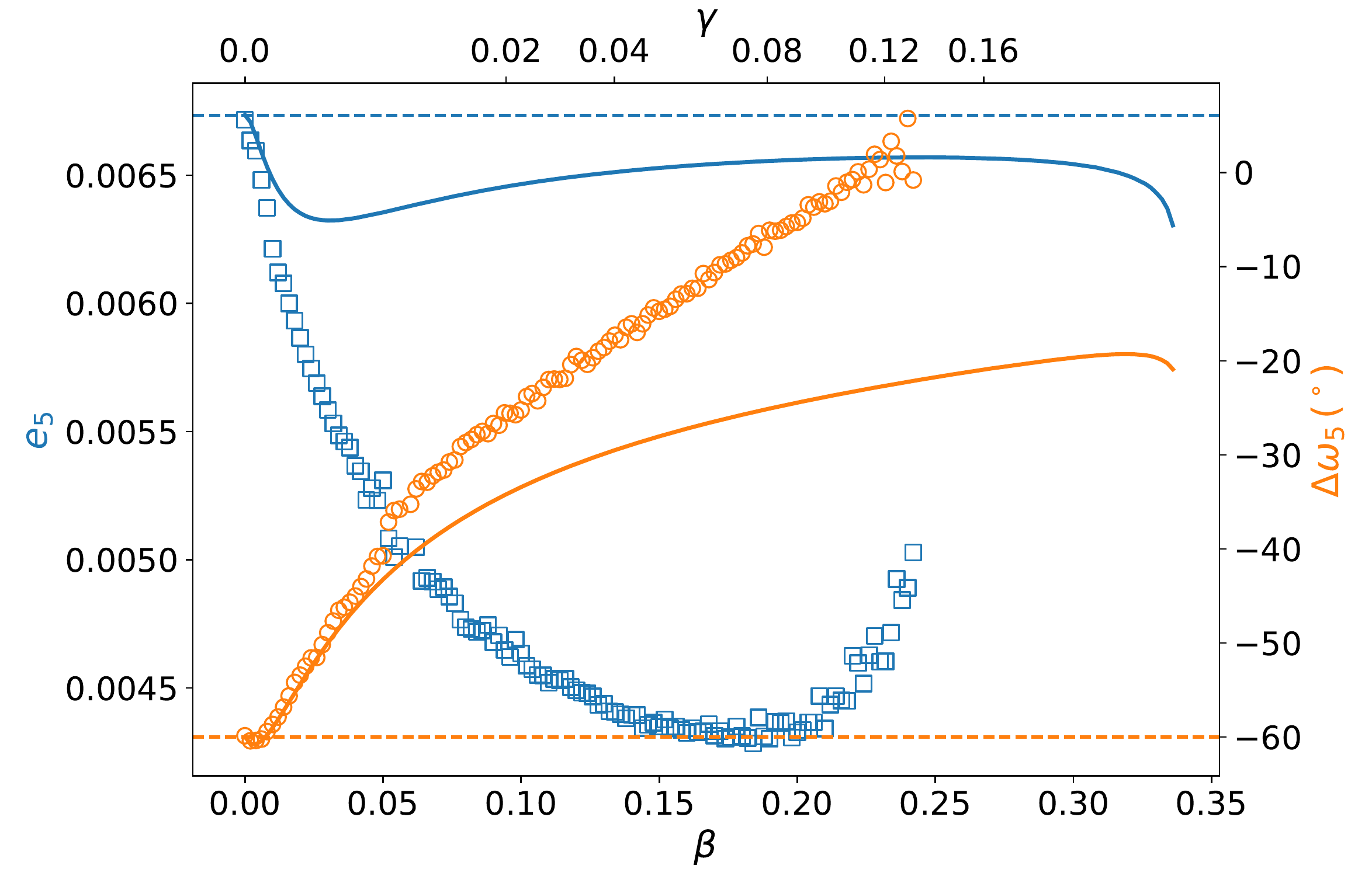}}
                \caption{Same as Fig.~\ref{fig:eome}, but dots indicate the model including the Lorentz force given a surface potential of 10 V.}
\label{fig:eomeqm}
        \end{figure}

We show in Fig.~\ref{fig:eomeqm} the apsidal difference from Venus ($\Delta\omega_k$) and the eccentricity ($e_k$) of the libration centers in a charged problem (shown in dots) obtained with the same method mentioned in Sec.~\ref{subsec:ns}. Apparently, they differ from the results in the uncharged problem (shown in curves) under the effect of the Lorentz force. We conclude that the Lorentz force tends to increase $\Delta\omega_k$ but decreases $e_k$ for both $L_4$ and $L_5$ in most situations. For $L_4$, $\gamma$ is limited to $3\times10^{-4}$ due to the maximum value of $\beta$ $0.01135$. Therefore the perturbation from the interplanetary field is small enough for us to determine $(\Delta\omega_k,e_k)$ precisely. Conversely, the orbits in $L_5$ with $\beta>0.05$ are much more chaotic. The lifespans of such orbits are reduced, and the libration centers cannot be located precisely, especially for $\beta>0.244$, for which $(\Delta\omega_k,e_k)$ cannot be determined because the lifespans are too short. However, we can still demonstrate that the Lorentz force can greatly change $e_k$ and $\Delta\omega_k$. The maximum differences could be up to 0.0022 and $25^\circ$ for $e_k$ and $\Delta\omega_k$, respectively.

Because many secular resonances control the dynamical behavior of particles in 1:1 MMR with Venus, the eccentricity and apsidal difference could experience complicated secular evolution. We therefore not show the plot of $(\Delta\omega_k,e_k)$ under the effect of the planetary perturbations.

\begin{figure}
        \centering
    \resizebox{\hsize}{!}{\includegraphics{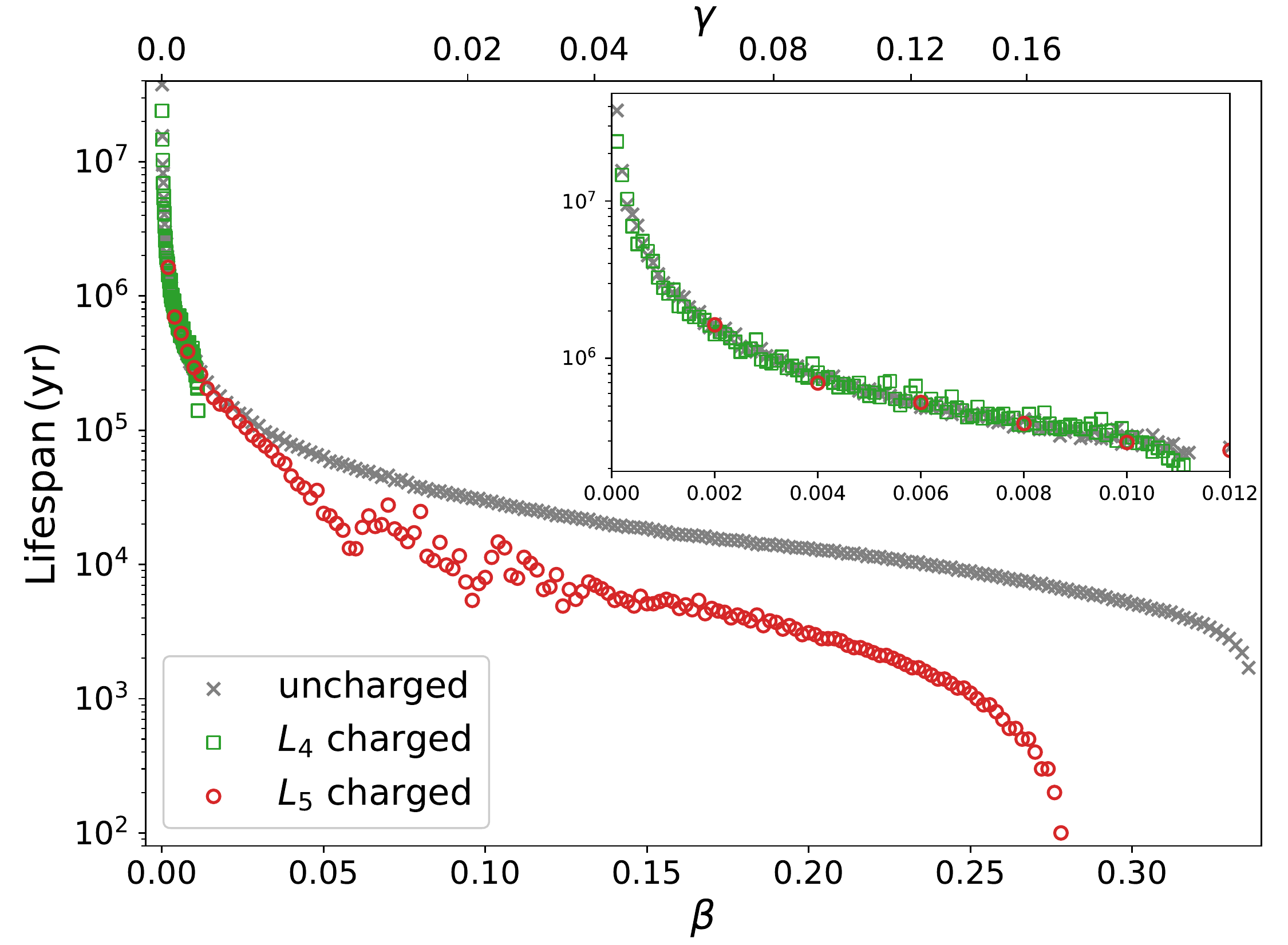}}
                \caption{Lifespans of dust particles in the libration centers of $L_4$ and $L_5$ in the charged problem. A window is inserted to zoom in the region of $0<\beta<0.012$. The gray dots indicate the lifespans in the uncharged problem that we showed in Fig.~\ref{fig:lf}.} \label{fig:lfqm}
        \end{figure}

The lifespans of charged particles in the libration centers again confirm that the perturbation from the interplanetary magnetic field further destabilizes the co-orbital region. Considering the uncertainties in determining the lifespans, we find a significant decrease in lifespans for $\beta>0.0105$ for both $L_4$ and $L_5$. The lifespan difference is about $10^5$ yr for $\beta=0.01135$, although the Lorentz force is smaller for smaller $\gamma$, but the longer lifespan makes it easier to accumulate the effect. On the other hand, the Lorentz force can also significantly reduce the stability within the short lifespan of large dust particles. The planetary perturbations can only cause a very limited difference to the lifespans compared to the Lorentz force. The most important dynamical effects caused by other planets on the co-orbital region are secular resonances \citepads[see, e.g.,]{2002MNRAS.334..241B,2009MNRAS.398.1217Z,2011MNRAS.410.1849Z,2012A&A...541A.127D,2019A&A...622A..97Z,2020A&A...633A.153Z}, which need a long timescale to be effective. Before this, however, dust particles are very likely to have escaped from the 1:1 MMR as a result of nongravitational effects.

\subsection{Parameter space study}

In this section we investigate the effect of charge on the full parameter space $(\beta,\gamma)$ with $0\leq\beta\leq0.01135$ for $L_4$ and $0\leq\beta\leq0.33865$ for $L_5$ and $\gamma$ derived from Eq. (\ref{eqn:bqm}), with $0\leq U\leq10$ Volt (for $L_4$ and $L_5$) with the aim to provide the domain in $\beta\times\gamma$ for which charge cannot be neglected in dynamical studies of dust in the Solar System. The results are based on the ERTBP with the standard Parker model for the interplanetary magnetic field. The initial conditions for the numerical simulations are as follows: for fixed pairs of parameters ($\beta$, $\gamma$), we chose initial conditions centered at the equilibria of $L_4$ and $L_5$ , taken from our previous analysis, together with initial conditions with slight deviations of $\delta a=\pm10^{-3}$ (normalized units), $\delta \sigma=\pm1^\circ$, $\delta\Delta\omega=\pm1^\circ$, or $\delta e=\pm e_{k}$ (with equilibrium value $e_k$ for $L_4$ and $L_5$, respectively). This gives nine initial conditions in total for each pair $(\beta, \gamma)$, but this is sufficient to investigate the effect of the charge parameter $\gamma$ on i) the time of temporary capture and ii) the maximum libration amplitudes compared to the uncharged case defined with $\gamma=0$. Let $T_\gamma$ be the time of temporary capture for a fixed pair of values ($\beta$, $\gamma$) normalized by the time of temporary capture for the same $\beta$, but with $\gamma=0$. 

\begin{figure}
        \centering
    \resizebox{\hsize}{!}{\includegraphics{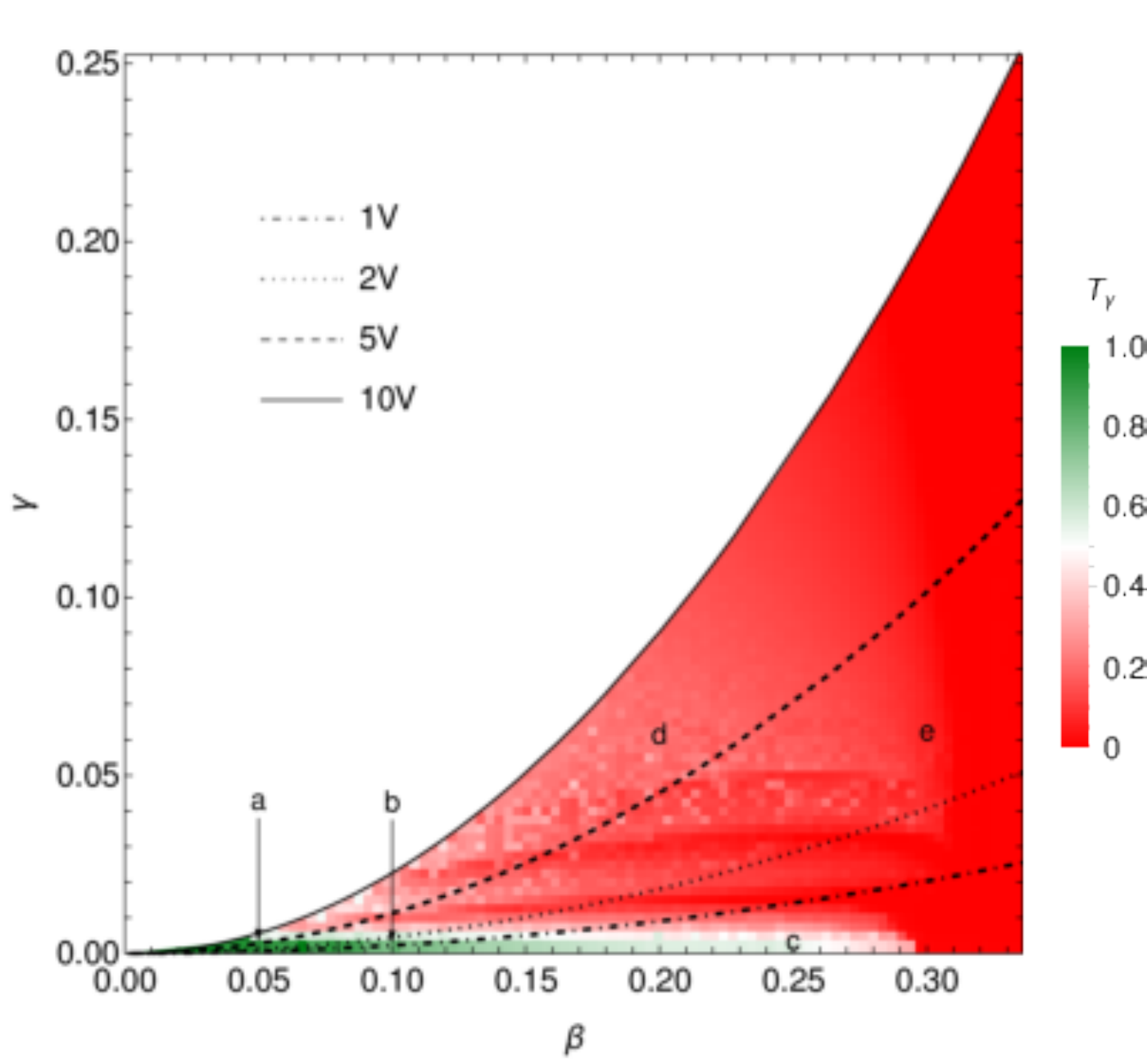}}
                \caption{Effect of charge-to-mass ratio $\gamma$ on stability (i.e., capture) time 
                $T_\gamma$ in the parameter space $(\beta\times\gamma)$.}
\label{fig:bega5}
        \end{figure}

\begin{figure}
        \centering
        \resizebox{\hsize}{!}{\includegraphics{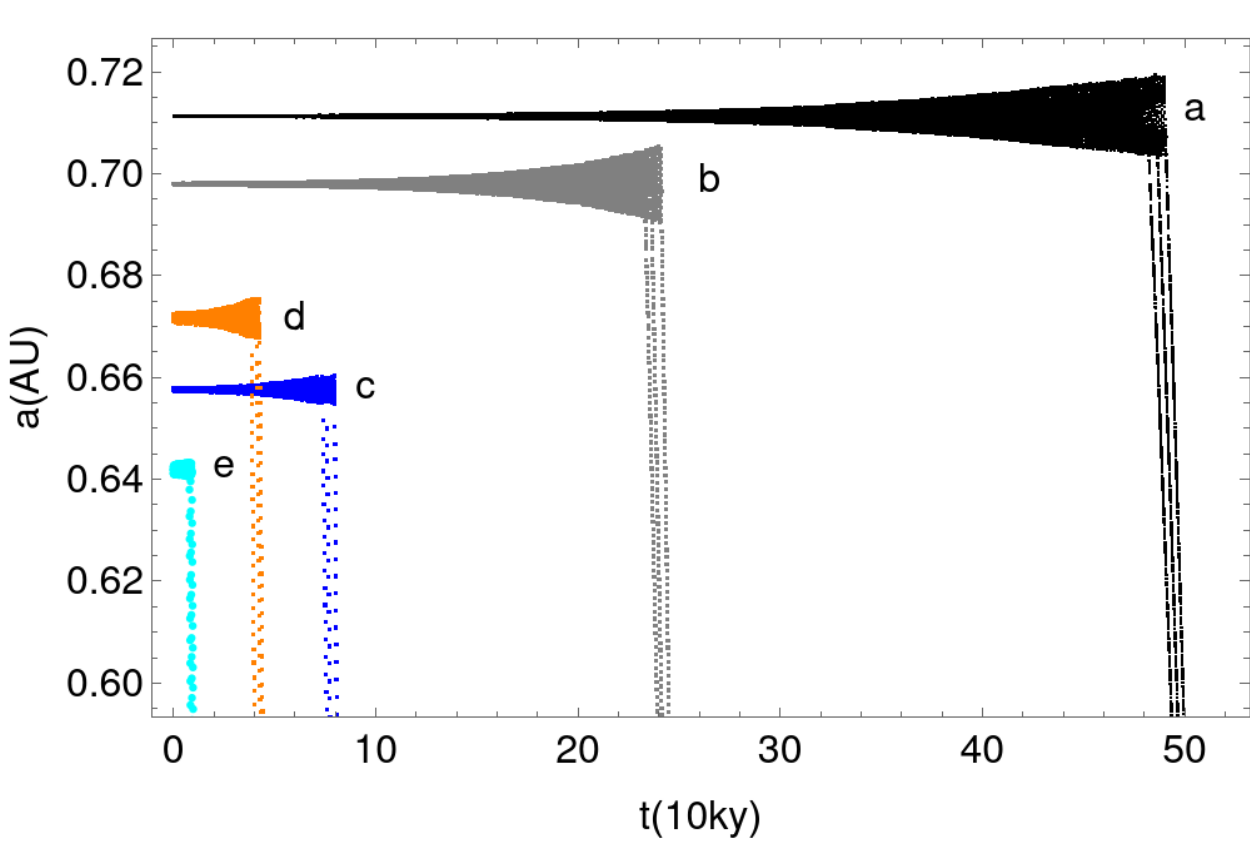}}
                \caption{Effect of different charge-to-mass ratios on the stability time $T_\gamma$. 
                Parameters for orbits a--e are taken from Fig.~\ref{fig:bega5} with initial conditions
                at the libration point $L_5$.}
\label{fig:a-e}
        \end{figure}

The results for the case $L_5$ are shown in Fig.~\ref{fig:bega5}, where we report the value of $T_\gamma$ with $0\leq T_\gamma\leq 1$ in the parameter space $(\beta, \gamma)$. As expected, for lower values of $\gamma,$ we find $T_\gamma>0.5$ up to $\beta\simeq0.3$ with $T_\gamma$ close to one for small $\beta$ and decreasing with increasing $\beta$. We also provide test cases within the green region in Fig.~\ref{fig:a-e} with three orbits each for values $(\beta,\gamma)$ in the dark green region (\textrm{a}), the light green region (\textrm{b}), and in the white region (\textrm{c}). We note that even low values of $\gamma$ are sufficient to obtain $T_\gamma\simeq0$ for $\beta>0.3$. With increasing values of $\gamma,$ we find decreasing $T_\gamma$ for the full range of $\beta\leq 0.3$ up to the threshold $\gamma\simeq0.01$ where $T_\gamma<0.5$. Within the light red regime of Fig.~\ref{fig:bega5} we still find capture,  with much shorter capture times, while in the dark red regions the perturbations due to charge dominate the other effects. For comparison we also provide orbits within the red regions (\textrm{d}, \textrm{e}) in Fig.~\ref{fig:a-e}. We clearly see the reduction of the capture time caused by charge and perturbations of the interplanetary magnetic field.  We conclude that the effects of the interplanetary magnetic field further destabilize motion in 1:1 resonance with Venus compared with vanishing charge-to-mass ratios in case of $L_5$.

The situation is somewhat different for the leading Lagrange point $L_4$. Already the standard nongravitational effect, that is, the P-R-S effect, destabilizes the orbits of captured dust in the uncharged case, and no capture persists for values with $\beta>0.01135$. For surface potentials $U\leq10$ V the effect of $\gamma$ is therefore not sufficiently large to act on the dynamics, that is, the capture processes. A similar study in the case of $L_4$ reveals that in the full parameter space $(\beta, \gamma),$ we find $T_\gamma\simeq1$ for all parameters and initial conditions (however, charge, i.e., the effect of the interplanetary magnetic field, still affects the maximum libration width; this is not shown here).


\section{Summary and conclusions}
\label{sec:con}

We investigated the role of nongravitational effects on the dynamics of charged micron-sized dust grains in the co-orbital region of Venus. Our study includes the investigation of planetary perturbations, solar radiation pressure, the Poynting-Robertson effect, and solar wind drag (P-R-S effect). In addition, we performed a detailed study of the effect of the Lorentz force as a result of the interaction with the interplanetary magnetic field (Parker spiral model), which is usually neglected in this type of studies \citepads[e.g.,][]{1994Icar..110..239B,1995Icar..113..403L,2012LPICo1667.6201J,2015Icar..250..249L}.

Our results are based on analytical estimates in the framework of simplified dynamical models, that is, the circular restricted three-body problem, and numerical studies in the elliptic restricted three-body problem, and the planetary problem including all nongravitational perturbations. We used a Fortran code that has been used in \citetads{2019CeMDA.131...49L} as well as SyMBA \citepads{1998AJ....116.2067D,2000AJ....120.2117L}, which we modified to include the additional perturbations, that is, the Lorentz form term.

We mainly focused on studying the position of the triangular Lagrange points $L_3$ (collinear), $L_4$ (leading), and $L_5$ (trailing) that define the geometry of the co-orbital regime of motions. We provided a detailed analysis of i) the shift of the Lagrange points and ii) the time of temporary capture of charged and uncharged dust grains in the co-orbital regime of motions, both due to the nongravitational effects. The shift and capture time strongly depend on the area-to-mass ratio in the dust grains, parameterized by the parameter $\beta$ and the charge-to-mass ratio $q/m$ parameterized by the parameter $\gamma$. 

As expected, we find that in the uncharged problem ($\gamma=0$), the shift due to solar radiation pressure is symmetric with respect to $L_4$ and $L_5$, while the P-R-S effect acts asymmetrically on the location of $L_4$ and $L_5$ . This phenomenon has previously been investigated in case of Jupiter \citepads{2015Icar..250..249L}. We derived an analytical formula to predict the shift for uncharged dust grains, based on the circular restricted three-body problem following the line of argumentation in \citetads{1980ApJ...238..337S}. We generalized the results to higher orders in the small parameters and also performed a linear stability analysis that provides an order-of-magnitude estimate of the e-folding time. We confirmed that the nongravitational effects render the equilibria unstable.

These results serve as the basis for finding good parameters and initial conditions for massive numerical simulations in the framework of the inclined elliptic restricted three-body problem and the full planetary problem. The aim is to find orbits with minimum libration amplitudes close to $L_4$ and $L_5$ that define the geometry of the co-orbital regime of motion also in these more realistic dynamical models. Denoting by $a_k$, $\sigma_k$, $e_k$, and $\Delta\omega_k$, with $k=4,5$ the semimajor axes, resonant angle, orbital eccentricity, and difference in argument of perihelion (with respect to Venus) with minimum libration amplitudes, close to the triangular Lagrange points, we provided their dependence on $\beta$ as well as $\gamma\neq0$. We find that nongravitational perturbations destroy librational orbits in the vicinity of $L_4$  already at $\beta\simeq0.01135$ (collision of $L_3$ with $L_4$), while librational orbits survive close to $L_5$ up to $\beta\simeq0.33865$ (collision of $L_1$ with $L_5$). This difference, caused by the P-R-S effect, results in an asymmetry in the the shift of $L_4$ and $L_5$ (with respect to the pure gravitational problem) and in different extensions of the regimes of librational motion of dust close to $L_4$ and $L_5$. The tadpole region expands with $\beta$ for $L_5$ almost linearly, but for $L_4$ it shrinks. For the width of the co-orbital region, which also contains the horseshoe orbits, we can always find some turning point ($\beta\sim0.07$) where the width is the largest.

The interaction of charged dust particles ($\gamma\neq0$) with the interplanetary magnetic field contributes to the shift of the tadpole points as well, but to a much lesser extent. However, for large enough $\gamma,$ we also find a notable effect that is comparable to the effect of the planetary perturbations in the uncharged case. The role of charge is also more important for the trailing Lagrange point $L_5$ that ensures librational motion up to much higher values in $\beta$ that correspond to much smaller particles with much higher charge-to-mass ratios for fixed values of density and surface potential. The Lorentz force therefore also affects the libration points in the co-orbital regime of motions in an asymmetric way, supporting librational motion of tiny dust grains close to $L_5$. Even more importantly, the Lorentz force term will act on the orbital planes of motions. In the uncharged problem, the dust grains stay within the orbital plane of Venus, neglecting the additional perturbations of the planets. However, the normal force component due to the Lorentz force will bring the dust particles out of their initial plane with latitudinal libration amplitudes up to several degrees already in the circular restricted problem. This phenomenon has been studied in full detail in \citetads{2019CeMDA.131...49L} and is related to the deviation of the magnetic axis of the sun from the axis of orbital angular momentum. Deviations from the orbital plane of Venus may reach several degrees. An additional important effect of the interplanetary magnetic field is the strong effect on the capture time of particles in the co-orbital regime of Venus.

It is commonly accepted that the orbital lifespan of (uncharged) dust grains in the Solar System is limited by the Poynting-Robertson effect. Outside of MMR with the planets, the effect leads to circularization and shrinking in semimajor axes, bringing the dust grains toward the Sun on spiral-like motions \citepads{1979Icar...40....1B}. However, the net effect that leads to the decrease in semimajor axis can vanish in presence of outer and co-orbital resonance with a planet, with the result of temporary capture of dust at specific distances from the Sun. This capture may strongly extend the lifespan of dust grains in the Solar System. In a detailed study of time of temporary captures of charged and uncharged dust grains in the co-orbital regime with Venus, we find a strong dependence on the area-to-mass ratio, related to parameter $\beta$, that is already known. However, we also see a strong dependence of the capture time on the parameter $\gamma;$  the charge-to-mass ratio of the dust particle as well. The reason is found in the faster increase of the libration amplitudes of charged dust in comparison to uncharged one. While for $0\leq\beta\leq0.3$ the capture time (in the vicinity of $L_5$) is about the same for $\gamma=0$ and $\gamma\leq0.01$, even low values of $\gamma\neq0$ may lead to release from co-orbital resonance at very short times. For higher charge-to-mass ratios $\gamma>0.01$ the time of temporary capture may be reduced by a factor 2-10 depending on the actual value of $\gamma$. Finally, we find that the effect of the interplanetary magnetic field on capture times close to $L_4$ and $L_5$ is asymmetric as well; it is much more important in the vicinity of $L_5$.

We remark that our study is based on a simplified model of the interplanetary magnetic field and assuming time-independent (equilibrium) charge-to-mass ratios of the dust grains. This is an oversimplification of the real situation in our Solar System. Time-dependent effects on the polarity and magnitude of the interplanetary magnetic field and the role of local perturbations due to coronal mass ejections have not been investigated in the present study at all. Time-dependent charge due to changes in the local plasma environment in space may result in additional effects that we have not touched on either.

However, the motivation of this study was to provide details on the role of the mean interplanetary magnetic field on the dynamics that is usually neglected by other authors. It may also serve as a basis for future studies that include more realistic models of the heliosphere. To conclude, the role of charge and the interplanetary magnetic field on the orbital motion of dust in the Solar System has been shown to be important to better understand the distribution and timescales of dust in our Solar System.

\begin{acknowledgements}
        This work has been supported by the Austrian Science Fund (FWF) in the framework of the stand alone project P30542, National Natural Science Foundation of China (NSFC, Grants No. 11473016, No. 11933001) and National Key R\&D Program of China (2019YFA0706601). Lei Zhou acknowledges the financial support from the program of China Scholarship Council (No. 201906190106) for his visit in University of Vienna.
\end{acknowledgements}

  \bibliographystyle{aa-note} 
  \bibliography{Venus_Dust.bib}

\begin{appendix}

        \section{Location shift of co-orbital resonances in the Solar System}

\subsection{Location of triangular Lagrangian points}\label{apd:sol}

To the first order in $\gamma_c$, Equations (\ref{eqn:r12}) and (\ref{eqn:r2}) can be simplified as
\begin{equation}\label{eqn:r12fod}
        \left[\frac{\gamma_c\,(1+s_w)}{2}+(1-\mu')\,y\right]r_1^3=\delta^3\,(1-\mu')\,y,
\end{equation}
\begin{equation}\label{eqn:r2fod}
        r_2 = 1+\frac{\gamma_c\,(1+s_w)}{3\,\mu'\,y}.
\end{equation}
Assuming that $y=y_0+\epsilon\,\gamma_c+o(\gamma_c^2),$ where $\epsilon$ is the coefficient of the first-order term in $\gamma_c$, we can obtain $r_1$ and $r_2$ in the form
\begin{eqnarray}\label{eqn:fodsol1}
        r_1^{(1)}&=&\delta\,\left[1-\frac{\gamma_c\,(1+s_w)}{6\,(1-\mu')\,y_0}\right],\\\label{eqn:fodsol2}
        r_2^{(1)}&=&1+\frac{\gamma_c\,(1+s_w)}{3\,\mu'\,y_0}.
\end{eqnarray}
Substituting $r_1$ and $r_2$ with Eqs. (\ref{eqn:fodsol1}) and (\ref{eqn:fodsol2}) in Eq. (\ref{eqn:xpu}), we obtain
\begin{equation}\label{eqn:xfod}
        x^{(1)} = x_0\,\left[1-\frac{(1-\mu')+\mu'\,\delta^2/2}{3\,(1-\mu')\,\mu'\,x_0\,y_0}(1+s_w)\,\gamma_c\right].
\end{equation}
We note that Eq. (\ref{eqn:xfod}) is a corrected version of Eq.~19 in \citetads{1980ApJ...238..337S}, and the first equation in Eq.~4 of \citetads{1995Icar..113..403L} because both of them missed $(1-\mu')$, which is the mass fraction of the star, in the denominator. This only causes a negligible difference because the planet mass is far lower than the stellar mass, however, and thus $(1-\mu')\approx1$. Recalling $r_2^2=(x+\mu'-1)^2+y^2$ and making use of Eqs. (\ref{eqn:fodsol2}) and (\ref{eqn:xfod}), we can obtain $y$ to first order in $\gamma_c$:
\begin{equation}\label{eqn:yfod}
        y^{(1)} = y_0\,\left\{1+\frac{\delta^2\,\left[(1-\mu')-\mu'\,\left(1-\delta^2/2\right)\right]}{6\,(1-\mu')\,\mu'\,y_0^3}(1+s_w)\,\gamma_c\right\}.
\end{equation}
Equaiton (\ref{eqn:yfod}) agrees perfectly with Eq.~20 in \citetads{1980ApJ...238..337S} and the second equation in \citetads{1995Icar..113..403L}.

Similarly, assuming $y=y_0+\epsilon_1\,\gamma_c+\epsilon_2\,\gamma_c^2+o(\gamma_c^3)$ with coefficients $\epsilon_{1,2}$ for the corresponding terms and making use of Eq. (\ref{eqn:yfod}), we can solve Eqs. (\ref{eqn:r12}) and (\ref{eqn:r2}) up to the second order in $\gamma_c$,
\begin{equation}\label{eqn:sodsolr1}
\begin{aligned}
        r_1^{(2)}=r_1^{(1)}+\frac{\gamma_c^2\,(1+s_w)^2\left[\delta^2\,(2 - \mu')-4(1 - \mu')\right]}{9 \left(4 - \delta^2\right)\,\delta\,(1 - \mu')^2\,\mu'\,y_0^2},
\end{aligned}
\end{equation}
\begin{equation}\label{eqn:sodsolr2}
        r_2^{(2)}=r_2^{(1)}-\frac{\gamma_c^2\,(1+s_w)^2\left[6-4\mu'-(2-\mu')\,\delta^2\right]}{9\,\left(4-\delta^2\right)\,(1-\mu')\,\mu'^2\,y_0^4},
\end{equation}
or, equivalently,
\begin{equation}\label{eqn:sodsolx}
\begin{aligned}
        x^{(2)} = & x^{(1)}-\gamma_c^2(1+s_w)^2\times \\
        & \frac{\delta^4{\mu'} ^2-4 \delta^2\,\left(2 {\mu'} ^2-4 \mu'+5\right)+16 \left({\mu'} ^2-5 \mu' +4\right)}{72 \left(4-\delta^2\right)(1-\mu' )^2\,{\mu'} ^2\,y_0^2}
\end{aligned}
\end{equation}
\begin{equation}\label{eqn:sodsoly}
\begin{aligned}
        y^{(2)} = y^{(1)}- & \frac{\gamma_c^2(1+s_w)^2}{144(1-{\mu'})^2\,{\mu'}^2\,y_0^5}\cdot \left\{{\delta^4 \left(5 {\mu'} ^2 y_0^2+4\right)+}\right. \\
        & \left.{2 \delta^2\left[\left(-11 {\mu'}^2 +8{\mu'} -10\right) y_0^2+4\right]+8 {\mu'} ^2 y_0^2 \left(2y_0^2-1\right)}\right\}
\end{aligned}
\end{equation}

Up to the third order in $\gamma_c$, we have
\begin{equation}\label{eqn:todsolr1}
\begin{aligned}
        r_1^{(3)}=r_1^{(2)}+ & \frac{\gamma_c^3\,(1+s_w)^3}{10368\,\delta (1 - \mu')^3 {\mu'} ^2 y_0^7}\times \\
        & \left[{3 \delta^{10} {\mu'} ^2-6 \delta^8 \left(5{\mu'} ^2-8 \mu' +10\right)+}\right. \\
        & \left.{48 \delta^6 \left({\mu'} ^2-4 \mu' +3\right)}\right. \\
    & \left.{-96\,\delta^4 \left(1 - \mu'^2\right)-192 \left(3 {\mu'} ^2-8 \mu' +5\right) y_0^4-}\right. \\
    & \left.{32 \delta^2 y_0^2 \left(12 \mu' +{\mu'} ^2 y_0^2-12\right)}\right],
\end{aligned}
\end{equation}
\begin{equation}\label{eqn:todsolr2}
        \begin{aligned}
        r_2^{(3)}=r_2^{(2)}+&\frac{\gamma_c^3\,(1+s_w)^3}{5184\,(1 - \mu')^2 {\mu'} ^3 y_0^9}\times \\
        & \left\{{3 \left(4-\delta^2\right) \delta^6\left[\left(4-\delta^2\right) \mu' -2\right]^2-}\right.\\
    & \left.{3 \delta^2 y_0^2 \left[{\delta^6 {\mu'} ^2+16 \delta^2(2 - \mu')^2 -32 \left(1 - \mu'^2\right)-}\right.}\right. \\
    & \left.{\left.{2 \delta^4 \left[\mu'\,(5 \mu' -8)+10\right]}\right]+896 (1 - \mu')^2 y_0^6+}\right. \\
    & \left.{192\delta^2 (1 - \mu') y_0^4 \left[\left(4-\delta^2\right) \mu' -2\right]}\right\},
\end{aligned}
\end{equation}
\begin{equation}\label{eqn:todsolx}
        \begin{aligned}
        x^{(3)} = x^{(2)} -&\frac{\gamma_c^3(1+s_w)^3}{10368 (1 - \mu')^3\,\mu'^3\,y_0^7}\times \\ 
   & \left\{{-3 \delta^{10} {\mu'} ^3+6 \delta^8 \mu'  \left(6 {\mu'} ^2-9 \mu' +10\right)+}\right. \\
   & \left.{12\delta^6 (1 - \mu')\left[-\mu'  (20 - 9 \mu')+10\right]+}\right. \\
   & \left.{64 (1 - \mu') \left[-\mu'  (35 - 4 \mu')+40\right] y_0^4+}\right. \\
   & \left.{48 \delta^4 \left[\mu'  \left(\mu' \left(y_0^2-10\right)+16\right)-6\right]+}\right. \\
   & \left.{32 \delta^2\left[{6 (1 - \mu')^2 (1 + \mu')+{\mu'} ^3 y_0^4-}\right.}\right. \\
   & \left.{\left.{6 (5 - \mu') (1 - \mu') y_0^2}\right]}\right\},
        \end{aligned}
\end{equation}
\begin{equation}\label{eqn:todsoly}
\begin{aligned}
        y^{(3)} = y^{(2)} - &\frac{\gamma_c^3(1+s_w)^3}{81 \left(4 - \delta^2\right)^4 \delta^4 (1 - \mu')^3 {\mu'} ^3}\times \\
   & \left[{\delta^8 {\mu'} ^3-2 \delta^6 \left(3 {\mu'} ^3+45 {\mu'} ^2-120 \mu' +80\right)+}\right. \\
   & \left.{8 \delta^4 \left(3 {\mu'} ^3+78 {\mu'} ^2-204 \mu' +115\right)-}\right.\\
   & \left.{32 \delta^2 \left(5 {\mu'} ^3+33 {\mu'} ^2-96 \mu' +56\right) + }\right. \\
   & \left.{384 \left({\mu'} ^3-4 \mu' +3\right)}\right].
\end{aligned}
\end{equation}

To further verify the approximation formulae, we also applied them to Jupiter, where $c_v\approx 22947$ (the figure is not shown here). We find a perfect consistency with a maximum deviation of 0.007$^\circ$ for $0<\beta<0.5$ for all formulae from first order to third order. The simplified formulae (see Eqs. (\ref{eqn:gensol}) and (\ref{eqn:simp})) are even less accurate than the first-order formulae in this case because the range of $\beta$ is larger than in the case of Venus, which causes large errors in the derivation of $p_{4,5}^{(m)}$ and $q_{4,5}^{(m)}$ by assigning $\beta=0$. In addition to the simplified formulae, the first- to third-order formulae are therefore still necessary in a study of nongravitational effects, especially for high values of $\beta$. Furthermore, for Jupiter, the resonant angle for $L_4$ $\sigma_4$ is also found to increase with $\beta,$ while the resonant angle for $L_5$ $\sigma_5$ decreases with $\beta$ as a result of dominating solar radiation pressure, as we mentioned in Sect.~\ref{subsec:loc}.

\subsection{Linear stability of triangular Lagrangian points}\label{apd:linearstab}

We considered a small displacement $(X, Y)$ from the equilibrium point $(X_0, Y_0)$ in the planar CRTBP including solar radiation pressure and the P-R-S effect. Substituting $(x,y)$ by $(X_0+X, Y_0+Y)$ in Eqs. (\ref{eqn:eomx}) and (\ref{eqn:eomy}) and expanding in a Taylor series up to the first order of $(X,Y,\dot{X},\dot{Y})$, we deduce the linearized equations\footnote{In this section, we denote $\gamma_c(1+s_w)$ as $\gamma_c'$}
\begin{equation}
\begin{aligned}\label{eqn:lsa}
        \ddot{X} +& \dot{X} \gamma_c' \left[\frac{(X_0+\mu')^2}{r_1^4}+\frac{1}{r_1^2}\right]+\dot{Y} \left[\frac{\gamma_c'  Y_0 (X_0+\mu')}{r_1^4}-2\right] = \\
    & X \left[{1-\frac{\delta ^3 (1-\mu')}{r_1^3}+\frac{3 \delta ^3 (1-\mu' ) (X_0+\mu')^2}{r_1^5}-\frac{2 \gamma_c'  Y_0 (X_0+\mu')}{r_1^4}-}\right. \\ 
    & \left.{\frac{\mu' }{r_2^3}+\frac{3 \mu' (X_0+\mu'-1)^2}{r_2^5}}\right]+ \\
    & Y \left[\frac{\gamma_c'}{r_1^2}+\frac{3 \delta ^3 (1-\mu' ) Y_0 (X_0+\mu')}{r_1^5}-\frac{2 \gamma_c' Y_0^2}{r_1^4}+\frac{3 \mu'  Y_0 (X_0+\mu'-1)}{r_2^5}\right],
\end{aligned}
\end{equation}
\begin{equation}
\begin{aligned}\label{eqn:lsb}
        \ddot{Y} + &\dot{X} \left[\frac{\gamma_c'  Y_0 (X_0+\mu')}{r_1^4}+2\right]+\dot{Y} \gamma_c'\left(\frac{Y_0^2}{r_1^4}+\frac{1}{r_1^2}\right) = \\
   & X \left[{-\frac{\gamma_c' }{r_1^2}+\frac{2 \gamma_c'  (X_0+\mu')^2}{r_1^4}+\frac{3 \delta ^3 (1-\mu' ) Y_0 (X_0+\mu')}{r_1^5}+}\right. \\
   & \left.{\frac{3 \mu'  Y_0 (X_0+\mu'-1)}{r_2^5}}\right]+\\
   & Y \left[{1-\frac{\delta ^3 (1-\mu')}{r_1^3}+\frac{2 \gamma_c'  Y_0 (X_0+\mu')}{r_1^4}+\frac{3 \delta ^3 (1-\mu' ) Y_0^2}{r_1^5}-}\right. \\
   & \left.{\frac{\mu'}{r_2^3}+\frac{3 \mu' Y_0^2}{r_2^5}}\right].
\end{aligned}
\end{equation}
We note that $r_1$ and $r_2$ above should be evaluated at the equilibrium points. The linear coefficients with respect to $X$, $Y$, $\dot{X}$, $\dot{Y}$ are denoted by $a_{ij}$ ($i,j=0...3$).

Up to second order in $\gamma_c$, we define the parameters $a_i\,(i=0..3)=\eta_1\,\gamma_c+\eta_2\,\gamma_c^2+o(\gamma_c^3)$ where $\eta_{1,2}$ are coefficients of the first and second terms (see Sec.~\ref{subsec:linearstab}). When $\gamma_c=0$, $a_i\,(i=0..3)=0,$ and thus Eq. (\ref{eqn:chareqn}) degenerates to Eq. (\ref{eqn:chareqns}) with solutions equal to $\lambda_0$ (see Eq. (\ref{eqn:chareqnssol})). Therefore the solution to Eq. (\ref{eqn:chareqn}) can be written as $\lambda=\lambda_0\,(1+c_1+\mathrm{i}c_2)$, where $c_i\, (i=1,2)=\zeta_1\,\gamma_c+\zeta_2\,\gamma_c^2+o(\gamma_c^3)$ with coefficients $\zeta_{1,2}$ for corresponding terms. By substituting $a_i\,(i=0..3)$ and $\lambda$ in Eq. (\ref{eqn:chareqn}) by series in $\gamma_c$, we can derive solutions to $c_1$ and $c_2$,
\begin{equation}
\begin{aligned}
        c_1=\frac{\gamma_c'}{4 \lambda_r^4-2 \lambda_r^2}\cdot &\left\{{-\frac{3 \left[a_{21} \lambda_r^2+9b a_{01} (1-\mu') \mu' \right]^2}{2 \left(1-2 \lambda_r^2\right)^2}+ }\right. \\
        & \left.{\frac{\left[a_{21} \lambda_r^3+9 b a_{01} \lambda_r (1-\mu') \mu \right]^2}{\left(2 \lambda_r^2-4 \lambda_r^4\right)^2}+}\right.\\
    & \left.{\frac{a_{21} \left[a_{21} \lambda_r^2+9 b a_{01} (1-\mu') \mu \right]}{2 \lambda_r^2-1} +}\right. \\
    & \left.{\frac{3 a_{31} \lambda_r^2 \left(a_{31} \lambda_r^2-a_{11}\right)}{2-4\lambda_r^2}-\frac{\left(a_{11}-a_{31} \lambda_r^2\right)^2}{4 \left(1-2 \lambda_r^2\right)^2}+}\right. \\
    & \left.{\frac{3 \left(a_{11} \lambda_r-a_{31} \lambda_r^3\right)^2}{2 \left(1-2 \lambda_r^2\right)^2}+\frac{a_{11} \left(a_{11}-a_{31} \lambda_r^2\right)}{2-4 \lambda_r^2}+}\right. \\
    &\left.{9 b a_{01} (1-\mu') \mu +a_{21} \lambda_r^2-}\right. \\
    & \left.{9 b a_{02} {\mu'} ^2+9 b a_{02} \mu'+a_{22}\lambda_r^2}\right\},
\end{aligned}
\end{equation}
\begin{equation}
\begin{aligned}
        c_2=\frac{\gamma_c'}{2 \lambda_r \left(2 \lambda_r^2-1\right)^3}\cdot&\left[{a_{11} \left({-18 b \gamma_c'  a_{01} {\mu'} ^2+18 b \gamma_c'  a_{01} \mu' +}\right.}\right. \\
   & \left.{\left.{\gamma_c'  a_{21}-4 \lambda_r^4+4 \lambda_r^2-1}\right)+4 a_{31}\lambda_r^6+}\right. \\
   & \left.{9 b \gamma_c'  a_{01} a_{31} {\mu'} ^2-9 b \gamma_c' a_{01} a_{31} \mu' - }\right. \\
   & \left.{\gamma_c'  a_{12} \left(1-2 \lambda_r^2\right)^2+2 \gamma_c'  a_{21} a_{31} \lambda_r^4-}\right. \\
   & \left.{2 \gamma_c' a_{21} a_{31} \lambda_r^2-4 a_{31} \lambda_r^4+a_{31} \lambda_r^2+}\right. \\
   & \left.{4 \gamma_c'  a_{32} \lambda_r^6-4 \gamma_c'  a_{32} \lambda_r^4+\gamma_c'  a_{32} \lambda_r^2}\right].
\end{aligned}
\end{equation}

\end{appendix}

\end{document}